\documentclass[12pt]{article}
\pdfoutput=1
\usepackage{amsmath}
\usepackage{amssymb}
\usepackage[pdftex]{graphicx}
\usepackage{xcolor}
\numberwithin{equation}{section}
\begin{document}

\vspace*{-1.5cm}
\thispagestyle{empty}
\begin{flushright}
AEI-2010-172
\end{flushright}
\vspace*{2.5cm}

\begin{center}
{\Large
{\bf Boundary conditions in Toda theories and minimal models}}
\vspace{2.5cm}

{\large Stefan Fredenhagen$^{1}$}
\footnotetext[1]{{\tt E-mail: Stefan.Fredenhagen@aei.mpg.de}}

\vspace*{0.5cm}

Max-Planck-Institut f{\"u}r Gravitationsphysik,
Albert-Einstein-Institut\\
D-14424 Golm, Germany\\
\vspace*{3cm}

{\bf Abstract}
\end{center}

We show that the disc bulk one-point functions in a $sl (n)$ Toda conformal field
theory have a well-defined limit for the central charge $c=n-1$, and that
their limiting values can be obtained from a limit of bulk one-point
functions in the $W_{n}$ minimal models. This comparison leads to a
proposal for one-point functions for twisted boundary conditions in
Toda theory.

\newpage

\tableofcontents

\section{Introduction}

Conformal field theories (CFTs) in two dimensions play an important
role in string theory and statistical physics. Since the seminal paper
by Belavin, Polyakov and Zamolodchikov~\cite{Belavin:1984vu},
enormous progress has been made. In particular, for rational CFTs we
have obtained a solid understanding, both in the mathematical
structures and in the tools that allow to determine
correlation functions (for a recent overview see~\cite{Fuchs:2009iz}). The
non-rational theories, in particular those with a continuous spectrum
(usually called non-compact theories), are much less understood. This
is unfortunate given that such theories are of prime importance when
we discuss e.g.\ AdS/CFT correspondence, in which non-compact target spaces
necessarily arise, or when we want to consider cosmological
backgrounds in string theory.

Interestingly, rational CFTs and non-compact CFTs have some points of
contact. Some families of rational CFTs have a non-rational limit
theory with a continuous spectrum, and the properties of the limit
theory can be understood from the rational CFT data. The first example
of such a point of contact is the Runkel-Watts
theory~\cite{Runkel:2001ng} that arises as the limit of unitary
Virasoro minimal models at central charge $c=1$. This theory can also
be understood~\cite{Schomerus:2003vv,Fredenhagen:2004cj} as a limit of
the non-compact Liouville
CFT. A similar story relates the $N=1$ supersymmetric minimal models and the $N=1$
super Liouville theory~\cite{Fredenhagen:2007tk}. Note that the notion of
a limit of CFTs is not unique, see~\cite{Roggenkamp:2003qp} for a different approach.

Liouville theory is the prime example of a non-compact CFT, and by now
we have achieved a very good understanding of this theory: the bulk
three-point functions are
known~\cite{Dorn:1994xn,Zamolodchikov:1995aa}, boundary conditions
have been found~\cite{Fateev:2000ik,Teschner:2000md,Zamolodchikov:2001ah}, and the
corresponding boundary structure constants have been
determined~\cite{Hosomichi:2001xc,Ponsot:2001ng,Ponsot:2003ss}. Liouville
theory can be seen as the $sl (2)$ case of the class of $sl (n)$ Toda
CFTs. These theories have large chiral symmetry algebras, the $W_{n}$
algebras, which for $n=2$ is just the Virasoro algebra. Toda theories
are interesting objects to study -- on the one hand they are highly
non-trivial examples for non-compact CFTs, on the other hand their large
symmetry makes us hope that they are still tractable. CFTs with
$W_{n}$ algebras are likely to play a role for the duals of higher
spin gauge theories on three-dimensional AdS
backgrounds~\cite{Campoleoni:2010zq,Henneaux:2010xg,Gaberdiel:2010pz}. Furthermore,
Toda theories appear in a recently proposed relation between
four-dimensional $N=2$ supersymmetric gauge theories and
two-dimensional CFTs~\cite{Alday:2009aq,Wyllard:2009hg}.

Despite their importance and the recent interest, much less is known
on Toda theories than on Liouville theory. The three-point correlators on
the sphere are only known for a subset of the primary
fields~\cite{Fateev:2007ab,Fateev:2008bm}. Topological defects in these theories
have been constructed from modular data in~\cite{Drukker:2010jp}.
Recently, boundary conditions in $sl (n)$ Toda CFTs have been
investigated~\cite{Fateev:2010za}, and bulk one-point functions in the
presence of these boundary conditions have been determined.

In view of the relation between Liouville theory and the Virasoro
minimal modes, it is natural to ask whether one can obtain Toda theories
as a limit of a family of rational CFTs. As suggested
in~\cite{Fredenhagen:2007tk}, one expects that the $W_{n}$ minimal
models approach a limit theory at central charge $c=n-1$, which
coincides with $sl (n)$ Toda CFT at this value of the central charge.

Boundary conditions in $W_{n}$ minimal models are completely
understood in terms of the Cardy construction~\cite{Cardy:1989ir} or
twisted versions thereof. One can thus use the point of contact at $c=n-1$ between
the minimal models and Toda CFT to test and interpret the Toda boundary
conditions of~\cite{Fateev:2010za}, and to improve our understanding
of them. In particular, one might hope to better understand the
divergences in the annulus partition functions from the limit of the
boundary spectra of minimal model boundary conditions.
\smallskip

In this paper, the limits of $sl (n)$ Toda CFT and of $W_{n}$ minimal
models are analysed and compared. In section~2 we study the spectrum
of Toda theory, the bulk two-point function, and the bulk one-point
function in the boundary theories of~\cite{Fateev:2010za} in the limit
$c\to n-1$. Section~3 reviews the $W_{n}$-minimal models and their
untwisted and twisted boundary conditions. Then, in section~4, we
define a limit theory for these models at $c=n-1$, and obtain a
continuous spectrum by averaging over the discrete minimal model
fields. In the limit theory, we consider the bulk two-point function,
as well as one-point functions for different classes of boundary
conditions. Up to a change of normalisation of the fields we find
complete agreement with the Toda analysis. We conclude in section~6
with the observation that the limit of twisted boundary conditions in
the $W_{n}$ minimal models leads to a precise proposal for the
one-point function for twisted boundary conditions in Toda theory,
which had not been determined in~\cite{Fateev:2010za}. Two appendices
contain details on the limit of the spectrum of the minimal models,
and on untwisted and twisted modular S-matrices.

\section{Boundary Toda conformal field theory}

In this section, we shall first review the bulk Toda theory, and
discuss the spectrum and the two-point functions in the limit $c\to
n-1$. Then we shall analyse the limit of one-point functions in the
boundary theories of~\cite{Fateev:2010za}.

\subsection{Bulk Toda theory}

Before we come to the boundary theory, we want to review a few facts
about Toda conformal field theory as can be found e.g.\ in~\cite{Fateev:2007ab}.
The two-dimensional $sl (n)$ Toda conformal field theory is described by the
action
\begin{equation}
S=\int \left( \frac{1}{8\pi} (\partial_{a}\phi)^{2}
+\frac{(Q,\phi )}{4\pi }R+ \mu \sum_{j=1}^{n-1}e^{b
(e_{j},\phi)}\right)\sqrt{g}d^{2}x\ ,
\end{equation}
where the scalar field $\phi= (\phi_{1},\dotsc ,\phi_{n-1})$ lives in
the Cartan subalgebra, the $e_{j}$ are the simple roots of $sl(n)$, $b$ is a
dimensionless coupling constant, and $\mu$ is called the cosmological
constant. $R$ is the scalar curvature of the two-dimensional
background metric $g$, and $Q$ is a background charge that takes the
value
\begin{equation}
Q=\left(b+b^{-1} \right)\rho
\end{equation}
for a conformally invariant theory (here $\rho$ denotes the Weyl
vector of $sl (n)$). The central charge of this theory is
\begin{equation}\label{centralcharge}
c=n-1+12Q^{2} = (n-1) (1+n (n+1) (b+b^{-1})^{2})\ .
\end{equation}
In addition to the energy momentum tensor there are higher spin
currents in the theory that form the $W_{n}$ algebra.
The spinless primary fields of Toda CFT are given by the exponentials
\begin{equation}
V_{\alpha} = e^{(\alpha ,\phi)}\ ,
\end{equation}
they are labelled by a vector $\alpha$. The conformal weight of
$V_{\alpha}$ is given by
\begin{equation}\label{Todaconfweight}
h (\alpha) =\frac{(\alpha ,2Q-\alpha)}{2} \ .
\end{equation}
For the physical spectrum we have $\alpha =Q+ip$ with a real vector
$p$, and the conformal weights are non-negative real numbers.
The conformal weights and all representation
properties of $V_{\alpha}$ and $V_{Q+w (\alpha -Q)}$ are the same for
any Weyl transformation $w \in W$, and the corresponding fields
coincide up to a factor,
\begin{equation}\label{reflection}
V_{\alpha} (z) = R_{w} (\alpha) V_{Q+w (\alpha-Q)} (z) \ .
\end{equation}
The reflection amplitude that occurs here is given~\cite{Fateev:2001mj}
by
\begin{equation}\label{reflamplitude}
R_{w} (\alpha) = \frac{A (Q+w (\alpha -Q))}{A (\alpha)} \ ,
\end{equation}
with 
\begin{equation}\label{reflfactor}
A (\alpha) = \left(\pi \mu \gamma (b^{2}) \right)^{b^{-1}(\alpha
-Q,\rho )} \prod_{e>0} \frac{2\pi b^{-1} }{\Gamma (b (\alpha -Q,e)) \Gamma (1+b^{-1}
(\alpha -Q,e))} \ ,
\end{equation}
where we take the product over all positive roots. We normalise the two-point
correlation functions to be
\begin{equation}
\langle V_{\alpha_{1}} (z_{1}) V_{\alpha_{2}} (z_{2})\rangle = \sum_{w\in W}
\delta (p_{1}+w (p_{2})) \frac{A (2Q-\alpha_{1})}{A (\alpha_{2})} 
|z_{1}-z_{2}|^{-4h (\alpha_{1})} \ ,
\end{equation}
where $\alpha_{j}=Q+ip_{j}$, and the delta distribution is defined with respect
to the standard metric on the weight space. Note that the sum over
the Weyl orbit is necessary to be consistent with the
identifications~\eqref{reflection} under Weyl transformations.  If we
choose to label fields only by their representatives $V_{\alpha}$ with
$p=-i (\alpha-Q)$ being in the interior of the fundamental Weyl chamber, the
two-point function reads
\begin{equation}\label{Todatwopoint}
\langle V_{\alpha_{1}} (z_{1})V_{\alpha_{2}} (z_{2})\rangle 
= \delta (p_{1}-p_{2}^{+})\frac{A
(2Q-\alpha_{1})}{A (\alpha_{2})} |z_{1}-z_{2}|^{-4h (\alpha_{1})} \ , 
\end{equation}
where $p^{+}$ is the conjugate weight vector of $p$, $(p^{+})_{j}= (p)_{n-j}$
for $j=1,\dotsc ,n-1$. Here, the $(p)_{j}$ are the coefficients of $p$
with respect to the basis of fundamental weights $\omega_{j}$,
$p=\sum_{j=1}^{n-1} (p)_{j}\omega_{j}$.

\subsection{Taking the limit}

We are interested in the connection between Toda CFTs and the
corresponding minimal models. The minimal models of the $W_{n}$
algebra all have central charge smaller than the rank $n-1$, with
$n-1$ being the supremum of all central charges. To make contact to
these models we therefore would like to take the limit of the $sl (n)$
Toda CFT to central charge $n-1$. From~\eqref{centralcharge} we see
that this value of the central charge is reached for $b=i$ (or
$b=-i$). Now it is a priori not clear whether such a continuation of
Toda CFT makes sense. However, it was shown in~\cite{Schomerus:2003vv}
that in Liouville theory the bulk correlation functions have a
well-defined limit for $b\to i$. The same holds true for the
correlators in $N=1$ supersymmetric Liouville
theory~\cite{Fredenhagen:2007tk}. In~\cite{Fredenhagen:2004cj} it was
shown for Liouville theory that also the boundary conditions have a well-defined limit
as $b\to i$. This suggests to evaluate this limit also for Toda CFTs.

\noindent We follow the strategy of~\cite{Schomerus:2003vv,Fredenhagen:2004cj}
and set 
\begin{equation}
\alpha =Q+ip \ ,
\end{equation}
where we keep $p$ constant in the limit. The factor $A(\alpha )$
defined in~\eqref{reflfactor} assumes the limit
\begin{equation}
\tilde{A} (p) = (\pi \mu_{\text{ren}})^{(p,\rho)} \prod_{e>0}
\left(2i \sin \pi (e,p) \right) \ .
\end{equation}
Here we have introduced the renormalised cosmological constant
$\mu_{\text{ren}}$. The reflection amplitude~\eqref{reflamplitude} is then given by
\begin{equation}
R_{w} (p) = \frac{\tilde{A} (w (p))}{\tilde{A} (p)}\ .
\end{equation}
The two-point function~\eqref{Todatwopoint} of fields $V_{ip}$ with $p$ in the fundamental
Weyl chamber reads in the limit
\begin{align}
\langle V_{ip_{1}} (z)V_{ip_{2}} (w)\rangle & =
\delta (p_{1}-p_{2}^{+}) \frac{\tilde{A} (-p_{1})}{\tilde{A} (p_{2})}
|z-w|^{-4h_{p_{1}}} \\
& = \delta (p_{1}-p_{2}^{+}) \left(\pi \mu_{\text{ren}} \right)^{-
(p_{1}+p_{2},\rho)} (-1)^{\frac{n (n-1)}{2}} |z-w|^{-4h_{p_{1}}} \ ,
\label{Todatwopointlimit}
\end{align}
where the limit of the conformal weight~\eqref{Todaconfweight} is
\begin{equation}
h_{p} = \frac{1}{2}p^{2} \ .
\end{equation}
The exponent of the sign in~\eqref{Todatwopointlimit} is given by the
number of positive roots in $sl (n)$.

The three-point correlation functions in $sl (n)$ Toda CFT are not known
in general, therefore we do not know how to take the limit. On the
other hand, they are known for a restricted set of
fields~\cite{Fateev:2007ab,Fateev:2008bm}, and it would be interesting
to evaluate the limit of those and compare it to the minimal model
side.

\subsection{One-point functions}

Due to the conformal symmetry, the one-point function of a bulk field
$\Phi_{\alpha}$ on the complex upper half plane with a conformal
boundary condition at the real axis is given by
\begin{equation}
\langle \Phi_{\alpha}\rangle_{s} = U_{s} (\alpha)
\frac{1}{|z-\bar{z}|^{2h (\alpha)}} \ ,
\end{equation}
where $s$ labels the boundary condition, and
$h (\alpha) =\bar{h} (\alpha)$ is the conformal weight of the bulk
field $\Phi_{\alpha}$ (only fields with $h=\bar{h}$ can couple to a
conformal boundary condition). The coefficients $U_{s} (\alpha)$ characterise
the boundary theory labelled by $s$.

Conformal boundary conditions and their one-point coefficients $U^{T}$
for $sl (n)$ Toda CFTs have been
determined in~\cite{Fateev:2010za}. The computations are done
explicitly for $sl (3)$, but it is suggested that similar formulae
also hold for arbitrary $n$. The non-degenerate
boundary conditions of~\cite{Fateev:2010za} are labelled by a vector
$s$, and the corresponding one-point coefficients are given
by\footnote{Note that our $s$ is related to the parameter $s_{\text{FR}}$
of~\cite{Fateev:2010za} by $s_{\text{FR}}=-2\pi s$. Also we use a
slightly different normalisation of the fields.}
\begin{align}
U^{T}_{s} (\alpha) & = A (\alpha)^{-1} \sum_{w\in W} e^{-2\pi (w (s),\alpha -Q)}
\nonumber\\
&= \big[\pi \mu \gamma (b^{2}) \big]^{\frac{(\rho
,Q-\alpha)}{b}} \prod_{e>0} \frac{\Gamma \big(b (e,\alpha -Q) \big)
\Gamma \big(1+b^{-1} (e,\alpha -Q) \big)}{2\pi b^{-1}} 
\sum_{w\in W}e^{-2\pi (w (s),\alpha
-Q)} \ .
\label{nondeg-boundary}
\end{align}
We see immediately that the boundary condition only depends on the
Weyl orbit of $s$. Also, the one-point
function~\eqref{nondeg-boundary} has the expected
reflection property (see~\eqref{reflection}),
\begin{equation}
U^{T}_{s} (\alpha ) = R_{w} (\alpha) U^{T}_{s} (Q+w (\alpha -Q)) \ .
\end{equation}
To the boundary condition labelled by $s$, one can associate the
so-called boundary cosmological constants~\cite{Fateev:2010za},
\begin{equation}
\lambda_{i,\pm} = \chi_{\omega_{i}} (2\pi b^{\pm 1}s) \ ,
\end{equation}
where $\chi_{\omega_{i}}$ is the character of the representation with
highest weight vector the $i^{\text{th}}$ fundamental weight
$\omega_{i}$ of $sl (n)$, $i=1,\dotsc ,n-1$.

In addition to the non-degenerate boundary conditions that correspond
to $(n-1)$-dimensional branes, there are degenerate boundary
conditions, which are associated to lower-dimensional
branes.\footnote{The dimensionality can be deduced on the one hand
from the comparison with the classical analysis, on the other hand it
is related to the infrared divergence of the one-point function;
see~\cite{Fateev:2010za} for details.}  In~\cite{Fateev:2010za}, these
are described for $sl (3)$, but their results suggest a
straightforward generalisation to arbitrary $sl (n)$. In fact, from
the Cardy construction~\cite{Cardy:1989ir} we expect that boundary
conditions are labelled by representations of the $W_{n}$ algebra.  In
addition to the generic representations without singular vectors there
is a hierarchy of (partly) degenerate representations (see e.g.\
\cite{Drukker:2010jp}). Their structure suggests that the general
maximally symmetric untwisted boundary conditions are labelled by a
subgroup $W'\subset W$ of the Weyl group, a vector $\kappa$ that is
invariant under $W'$, and two dominant integral weights $\Omega
,\Omega'$. The subgroup $W'$ is not arbitrary, but it is generated by
reflections corresponding to a subset $\{\alpha_{i} \}_{i\in S}$ of the
simple roots of $sl (n)$ ($S$ being a subset of $\{1,\dotsc ,n-1
\}$). The subspace $V$ of weight vectors $v$ that are invariant under
$W'$ is then characterised by the condition $(v,\alpha_{i})=0$ for all
$i\in S$. If we consider a generic vector $v\in V$, i.e.\ one that
satisfies $(v,\alpha_{i})\not= 0$ for $i\not\in S$, then the subgroup
$W'$ equals the stabiliser group $W_{v}$ of $v$,
\begin{equation}
W' = W_{v} := \big\{ w\in W \, |\,  w (v) = v \big\} \ .
\end{equation}
On the other hand, any stabiliser group $W_{v}$ with $v$ in the
fundamental Weyl chamber is generated by the simple reflections along
those simple roots $\alpha_{i}$ that are orthogonal to $v$. Therefore we can
characterise the allowed subgroups $W'$ as stabiliser groups of
vectors in the fundamental Weyl chamber.

The coefficient of the bulk one-point function is given as a
sum over the coefficients $U^{T}_{s} (\alpha)$ of the non-degenerate
boundary conditions,
\begin{equation}\label{degenerateToda}
U^{T,W'}_{\kappa ,\Omega,\Omega '} (\alpha) = \sum_{w\in W'}
\epsilon (w) U^{T}_{s (\kappa ,\Omega ,\Omega ',w)} (\alpha) \ ,
\end{equation}
where 
\begin{equation}
s (\kappa ,\Omega ,\Omega ',w) = \kappa - i (b
(\Omega +\rho) + b^{-1} w(\Omega '+\rho))\ .
\end{equation}
Note that the boundary cosmological constants associated to the
different $s(\kappa ,\Omega ,\Omega ',w)$ are equal, they do not
depend on $w$. We have
\begin{equation}
\chi_{\omega_{j}} (2\pi bs (\kappa ,\Omega ,\Omega ',w)) =
\chi_{\omega_{j}} (2\pi b\kappa -2\pi ib^{2} (\Omega +\rho)-2\pi iw
(\Omega' +\rho)) \ ,
\end{equation}
and the right hand side is independent of $w$, because $w (\Omega
+\rho)$ differs from $\Omega +\rho$ by an element of the root lattice,
leading to a trivial phase in the character. Similarly
\begin{align}
\chi_{\omega_{j}} (2\pi b^{-1}s (\kappa ,\Omega ,\Omega ',w)) & =
\chi_{\omega_{j}} (2\pi b^{-1}\kappa -2\pi i (\Omega +\rho) -2\pi ib^{-2}w
(\Omega '+\rho ) ) \nonumber\\
& = \chi_{\omega_{j}} (2\pi b^{-1}w^{-1} (\kappa) -2\pi i w^{-1} (\Omega
+\rho) -2\pi ib^{-2} (\Omega '+\rho)) 
\end{align}
is independent of $w$, because $\kappa$ is invariant under $w\in
W'$. 

We call a boundary condition $m$-degenerate
if the subspace $V$ of vectors invariant under the group $W'$ is
$(n-1-m)$-dimensional. Note that the labels $\kappa ,\Omega ,\Omega '$
have in general a redundancy: the $W'$-invariant part
of $2\pi i ( b (\Omega +\rho)+b^{-1}w (\Omega '+\rho))$ can be
absorbed into $\kappa$. Therefore we have $(n-m-1)$ continuous parameters
to choose $\kappa$, and $2m$ discrete parameters labelling the
components of $\Omega ,\Omega '$ orthogonal to the invariant
subspace. If one requires that the
$W'$-invariant part of $s (\kappa ,\Omega ,\Omega ',w)$ is
real, then an $m$-degenerate boundary condition can be labelled by
the set $\mathbb{R}^{n-1-m}\times \mathbb{N}^{2m}$.

All the boundary conditions that we have described until now satisfy
trivial gluing conditions for the currents of the $W_{n}$-algebra. For
$n>2$, the algebra has an automorphism that is induced by an outer
automorphism of $sl (n)$. One can then also study twisted boundary
conditions, in which the currents satisfy gluing conditions that are
twisted by the automorphism. These boundary conditions have been
studied for $sl (3)$ in~\cite{Fateev:2010za}, but only in the light
asymptotic limit (where $b\to 0$),
for which bulk one-point and boundary two-point correlators have been
determined.
\smallskip

We now take the limit $b\to i$. We keep $ip=\alpha -Q$ fixed, and for
the non-degenerate boundary condition~\eqref{nondeg-boundary} the
limit $\tilde{U}^{T}$ of the one-point coefficient is
\begin{equation}\label{limitnondeg}
\tilde{U}^{T}_{s} (p) = \big[\pi \mu_{\text{ren}} \big]^{- (\rho ,p)}
\prod_{e>0} \left(2i \sin \pi (e,p)\right)^{-1} \sum_{w\in W} e^{-2\pi
i(w(s),p)}\ .
\end{equation}
The degenerate ones are then described by
\begin{equation}\label{limitdeg}
\tilde{U}^{T,W'}_{\kappa ,\Omega ,\Omega '} (p) 
= \sum_{w\in W'} \epsilon (w) \tilde{U}^{T}_{\kappa + (\Omega +\rho)
-w(\Omega '+\rho)} (p) \ .
\end{equation}
In the extreme case, $W'=W$, the only vector invariant under $W'$ is
the zero vector and we find the completely degenerate boundary
conditions
\begin{equation}\label{limitcompdeg}
\tilde{U}^{T}_{\Omega ,\Omega '} (p) \equiv \tilde{U}^{T,W}_{0,\Omega
,\Omega '} (p) = \sum_{w\in W} \epsilon (w) \tilde{U}^{T}_{(\Omega +\rho)
-w(\Omega '+\rho)} (p) \ .
\end{equation}
We will see later that not all of these are linearly independent (see
the discussion at the end of section~\ref{sec:minmodlimuntwist}). For
the twisted boundary conditions we cannot take the limit, as the
results of~\cite{Fateev:2010za} are only obtained in the limit $b\to
0$.

We are now going to analyse boundary conditions in minimal models,
where we want to reproduce~\eqref{limitnondeg} and~\eqref{limitdeg} in
the corresponding limit.

\section{Boundary conditions in minimal models}

In this section we review the $W_{n}$ minimal models and their
untwisted and twisted boundary conditions. 

\subsection{Bulk theory}

The $W_{n}$ minimal models can be obtained~\cite{Bais:1987zk} 
by a diagonal coset construction~\cite{Goddard:1986ee},
\begin{equation}
M_{n} (k) = \frac{sl (n)_{k}\oplus sl (n)_{1}}{sl (n)_{k+1}}\ .
\end{equation}
Their central charge is given by
\begin{equation}
c_{n} (k) = (n-1) \left(1-\frac{n (n+1)}{(k+n) (k+n+1)} \right)\ .
\end{equation}
The sectors of the theory are labelled by three integral dominant
weights $(\Lambda,\lambda;\Lambda')$ of the affine Lie algebras $sl
(n)_{k}, sl (n)_{1}$ and $sl (n)_{k+1}$, respectively. When we write
$\Lambda$ (and similarly for $\lambda$, $\Lambda '$), we think of it
as the finite part $\Lambda = (\Lambda_{1},\dotsc ,\Lambda_{n-1})$ of
an affine weight $(k -\Lambda_{1}-\dotsb
-\Lambda_{n-1},\Lambda_{1},\dotsc ,\Lambda_{n-1})$. These labels are
subject to the selection rule 
\begin{equation}\label{selrule}
\Lambda+\lambda-\Lambda'\in L_{R} \ ,
\end{equation}
where $L_{R}$ is the root lattice of $sl (n)$. This selection rule
determines $\lambda$ completely in terms of $\Lambda$ and $\Lambda'$,
and $\lambda$ can thus be omitted. In addition, some sectors have to be
identified according to the field identifications
\begin{equation}\label{fieldidentification}
(\Lambda;\Lambda') \sim (J\Lambda;J\Lambda') \ ,
\end{equation}
where $J$ is the generator of the $\mathbb{Z}_{n}$ simple current
groups (we denote the simple currents in $sl (n)_{k}$ and $sl
(n)_{k+1}$ by the same symbol). The action of the simple current $J$
on a weight $\Lambda$ at level $k$ is given by
\begin{equation}\label{simplecurrentaction}
J\Lambda = k\omega_{1} + w_{J}\lambda \ ,
\end{equation}
with the Weyl group element $w_{J}=s_{1}\dotsm s_{n-1}$
acting as
\begin{equation}\label{simplecurrentWeyl}
w_{J} (\Lambda_{1},\dotsc ,\Lambda_{n-1}) = (-\Lambda_{1}-\dotsb
-\Lambda_{n-1},\Lambda_{1},\dotsc ,\Lambda_{n-2}) \ .
\end{equation}
Here, $s_{i}$ are the Weyl reflections for the simple roots
$\alpha_{i}$, and $\omega_{i}$ are the fundamental weights.

\noindent The conformal weight $h_{\Lambda,\Lambda'}$ of a primary state with label
$(\Lambda;\Lambda')$ is given by
\begin{equation}\label{confweightminmod}
h_{\Lambda,\Lambda'} = \frac{1}{2t}\left(d_{\Lambda+\rho
,\Lambda'+\rho}^{2} - d_{\rho ,\rho}^{2} \right) \ ,
\end{equation}
where 
\begin{equation}
d_{v,v'} := v - t v'\quad ,\quad
t=\frac{k+n}{k+n+1} \ ,
\end{equation}
and $\rho$ is the Weyl vector. Under a field identification, the
vector $d_{\Lambda +\rho,\Lambda '+\rho}$ transforms by a Weyl transformation, 
\begin{align}
d_{J\Lambda +\rho ,J\Lambda' +\rho }&= (J\Lambda +\rho )  -t
(J\Lambda' +\rho ) \nonumber\\
& = \big( k\omega_{1}+w_{J}\Lambda +\rho \big) 
-t\big((k+1)\omega_{1}+w_{J}\Lambda' +\rho \big)\nonumber\\
& = \big((k+n)\omega_{1}+w_{J} (\Lambda +\rho) \big) - 
t \big((k+n+1)\omega_{1} + w_{J} (\Lambda '+\rho) \big)\nonumber\\
& = w_{J}\big(\Lambda +\rho - t (\Lambda '+\rho ) \big)  =
w_{J}d_{\Lambda +\rho,\Lambda'+\rho}\ .
\label{fieldidentreltoWeyl}
\end{align}
In going from the second to the third line we used that $\rho =
n\omega_{1}+w_{J} (\rho)$, which follows
from~\eqref{simplecurrentWeyl}.  We normalise the fields such that the
two-point correlator is given by
\begin{equation}\label{minmodtwopoint}
\langle \phi_{(\Lambda_{1};\Lambda_{1}')} (z)
\phi_{(\Lambda_{2};\Lambda_{2}')} (w)\rangle =
\delta_{(\Lambda_{1};\Lambda_{1}')
(\Lambda_{2}^{+};\Lambda_{2}'^{+})}|z-w|^{-4h_{\Lambda_{1},\Lambda
_{1}'}} \ . 
\end{equation}

\subsection{One-point functions for untwisted boundary conditions}

The maximally symmetric, untwisted boundary conditions in the $W_{n}$
minimal models are labelled by the same labels as the bulk fields, we
denote them by $(L;L')$. The one-point functions are then given by the
Cardy construction,
\begin{equation}
U^{M}_{(L;L')} (\Lambda ;\Lambda ') 
= \frac{S_{(\Lambda;\Lambda')
(L;L')}}{\sqrt{S_{(\Lambda;\Lambda') (0;0)}}}\ ,
\end{equation}
where $S$ is the modular S-matrix of the minimal model. It can be
expressed in terms of the modular S-matrix $S^{(n,k)}$ of the 
$sl(n)_{k}$ affine Lie algebra,
\begin{equation}
S_{(\Lambda,\lambda;\Lambda') (L,l;L')} = n
S^{(n,k)}_{\Lambda L}\overline{S^{(n,k+1)}_{\Lambda' L'}}
S^{(n,1)}_{\lambda l} \ ,
\end{equation}
where the bar over the S-matrix of the level $(k+1)$ part denotes
complex conjugation.
The S-matrix depends on $\Lambda$ and $\Lambda'$ only via the combination
$d_{\Lambda+\rho ,\Lambda'+\rho}$, and can be expressed as
(see~\eqref{app:cosetSfinal})
\begin{equation}
S_{(\Lambda;\Lambda') (L;L')} = \mathcal{N} \sum_{w,w' \in W} \epsilon
(ww') e^{-2\pi it^{-1}\big(w ( d_{\Lambda+\rho ,\Lambda'+\rho}),L+\rho
\big)} e^{2\pi i \big(w' (d_{\Lambda+\rho,\Lambda'+\rho}),L'+\rho
\big)} \ ,
\end{equation}
where 
\begin{equation}\label{defN}
\mathcal{N} = n^{-1/2} \left((k+n) (k+n+1) \right)^{- (n-1)/2} \ .
\end{equation}
In total we arrive at the bulk one-point coefficient
\begin{align}
U^{M}_{(L;L')} (\Lambda ;\Lambda ') & = A_{M}
(d_{\Lambda+\rho,\Lambda'+\rho})
\sum_{w\in W} \epsilon (w) e^{-2\pi it^{-1}\big(w (
d_{\Lambda+\rho ,\Lambda'+\rho}),L+\rho \big)} \nonumber\\
&\quad \times \sum_{w'\in W} \epsilon (w') e^{2\pi i \big(w'
(d_{\Lambda+\rho,\Lambda'+\rho}),L'+\rho \big)}\ ,
\label{mm-one-point}
\end{align}
with $A_{M} (d)$ given by
\begin{align}
A_{M} (d) & =\mathcal{N}^{1/2} \left( \sum_{w\in W} \epsilon
(w)e^{-2\pi it^{-1} (w (d),\rho)} \right)^{-1/2}\nonumber\\
& \quad \times \left( \sum_{w'\in W} \epsilon
(w')e^{2\pi i (w' (d),\rho)} \right)^{-1/2} \\
& = \mathcal{N}^{1/2} \prod_{e>0}\left(4\sin \left(\pi t^{-1} (e,d)
\right)\sin \left(\pi (e,d) \right) \right)^{-1/2} \ ,
\label{defAM}
\end{align}
where we used the Weyl denominator formula (see e.g.\
\cite{FrancescoCFT}). The product runs over the positive roots of $sl(n)$.

\subsection{One-point functions for twisted boundary conditions}
\label{sec:minmodtwisted}

For $n\geq 3$, the $sl (n)$ algebra has an outer automorphism $\omega$
coming from the reflection symmetry of the Dynkin diagram. Correspondingly,
the coset algebra also has an outer automorphism, and we can look for
boundary conditions that glue the right-moving and left-moving
currents with a twist given by this automorphism. 

The twisted boundary states for $SU(n)$ WZW models have
been constructed in~\cite{Birke:1999ik}. From this construction it is
straightforward to obtain the twisted boundary states in the associated
coset models~\cite{Ishikawa:2001zu,Fredenhagen:2003xf} (see
also~\cite{Caldeira:2004jy} for a discussion of boundary conditions in $W_{n}$
minimal models). 

The details of the construction depend on $n$ being even or odd. The
case of even $n$ is technically more complicated, because in the
standard construction of twisted coset boundary states one has to do a
fixed-point resolution. Although this can be solved in a straightforward
way, we shall concentrate here on the case of odd $n=2m+1$, where these
technical problems are absent.

The twisted boundary states of the $sl (2m+1)$ theories at level $k$
can be labelled by symmetric $sl (2m+1)$-weights $L= (L_{1},\dotsc
,L_{m},L_{m},\dotsc ,L_{1})$ with
$2\sum_{i=1}^{m}L_{i}\leq k$ (one should think of them as labels of
representations of the twisted affine Lie algebra
$A_{2m}^{(2)}$).  The coefficient of the bulk one-point
functions in the $SU (2m+1)$ WZW models for a twisted boundary
condition $L$ is
\begin{equation}
U^{M}_{\omega ,L} (\Lambda) =
\frac{\psi^{(n,k)}_{L \Lambda}}{\sqrt{S^{(n,k)}_{0\Lambda}}}
\delta_{\Lambda,\Lambda^{+}} \ .
\end{equation}
Here, $\psi$ is the twisted S-matrix (given in eq.\
\eqref{app:twistSfinal} in the appendix~\ref{sec:apptwistedS}). Only those
bulk-fields $\phi_{\Lambda}$ can couple that are invariant under the
automorphism, i.e.\ which are labelled by self-conjugate
representations $\Lambda=\Lambda^{+}$.

The twisted boundary conditions in the coset theory are then given by
three symmetric labels $L,\ell,L'$ at levels $k$, $1$ and $k+1$,
respectively. Note that $\ell$ can only take the value $\ell=
(0,\dotsc ,0)$, so that we can label the boundary states just by
$(L;L')$. There are no selection or identification rules for the
twisted coset boundary states.

\noindent The one-point coefficients are given by 
\begin{equation}
U^{M}_{\omega ,(L;L')} (\Lambda ,\lambda;\Lambda ')
= \frac{\psi^{(n,k)}_{L\Lambda}\psi^{(n,1)}_{0\lambda}
\overline{\psi^{(n,k+1)}_{L'\Lambda'}}}{\sqrt{n
S^{(n,k)}_{0\Lambda}S^{(n,1)}_{0\lambda}S^{(n,k+1)}_{0\Lambda '}}}
\delta_{(\Lambda ,\lambda;\Lambda '),
(\Lambda^{+},\lambda^{+};\Lambda'^{+})} \ .
\end{equation}
Only those bulk fields couple that are labelled by self-conjugate
representations. In the above formula it is understood that in the
field identification orbit of a self-conjugate coset representation
one chooses the unique representative that consists itself of
self-conjugate labels $\Lambda =\Lambda^{+}$, $\lambda =\lambda^{+}$
and $\Lambda '=\Lambda '^{+}$. Note that for odd $n$ the only
self-conjugate label $\lambda$ at level $1$ is 
$\lambda = (0,\dotsc ,0)$.
 
We want to rewrite the one-point functions in a way that is more
useful when we take the limit $k\to \infty$, namely we want to express
it such that the field labels $(\Lambda ;\Lambda ')$ only enter in the
combination $d_{\Lambda +\rho ,\Lambda'+\rho}$. To rewrite the product
of the twisted S-matrices we use eq.\ \eqref{app:cosettwistSfinal}
from the appendix, and we arrive at the following formula for the
one-point coefficient for self-conjugate bulk labels $\Lambda
=\Lambda^{+}$, $\Lambda '=\Lambda '^{+}$,
\begin{align}
U^{M}_{\omega ,(L;L')} (\Lambda;\Lambda ') & = n^{1/2} \left((k+n) (k+n+1)
\right)^{m/2} A_{M} (d_{\Lambda
+\rho ,\Lambda '+\rho}) \nonumber\\
& \quad \times \sum_{w,w'\in W^{\omega}} \epsilon (w)\epsilon
(w') e^{-\pi i\left(t^{-1}w
(L+\rho)-w' (L'+\rho),d_{\Lambda +\rho ,\Lambda '+\rho} \right)} \ , 
\label{twistedbc}
\end{align}
where $A_{M} (d)$ was given in~\eqref{defAM}. The subgroup
$W^{\omega}\subset W$ consists of all those Weyl transformations that
leave the subspace of symmetric weights invariant. 

\section{The limit of the minimal models}

\subsection{Bulk spectrum}
\label{sec:minmodlimitbulk}
As we have seen, the spectrum of a minimal model is labelled by two
integral, dominant weights $\Lambda ,\Lambda '$ of an affine $sl (n)$
algebra at level $k$ and $k+1$, respectively. Their conformal weight
(see eq.\ \eqref{confweightminmod}),
and actually all their higher W-charges are determined~\cite{Fateev:1987zh} by the
combination $d_{\Lambda+\rho,\Lambda '+\rho}$, or better by its Weyl
orbit. After rotating $d_{\Lambda+\rho,\Lambda'+\rho}$ to the 
fundamental Weyl chamber, these vectors
approach a uniform distribution in the whole fundamental Weyl chamber
in the limit $k\to \infty$. The spectrum hence becomes continuous in
this limit, and the primary states are labelled by vectors $d$ in the
fundamental Weyl chamber. Following the strategy
of~\cite{Runkel:2001ng} for the $sl (2)$ case, we want to
define fields $\phi_{d}$ in the limit theory as an average over fields
$\phi_{(\Lambda ;\Lambda ')}$ whose values $d_{\Lambda +\rho ,\Lambda
'+\rho}$ are close to $d$ in the limit (modulo a Weyl transformation).
As an approximation to the fields $\phi_{d}$, we introduce the
averaged fields
\begin{equation}\label{defaveragedfields}
\phi_{d}^{(\epsilon,k)} = \frac{1}{|N (d,\epsilon ,k)|} \sum_{(\Lambda
;\Lambda ')\in N (d,\epsilon ,k)} \phi_{(\Lambda ;\Lambda ')} \ ,
\end{equation}
where 
\begin{equation}\label{defofN}
N (d,\epsilon ,k) = \{(\Lambda;\Lambda ') :  \exists \ w\in
W\ \text{s.t.}\ |(w (d_{\Lambda +\rho ,\Lambda
'+\rho}))_{i}-d_{i} |<\epsilon/2 \ \ \text{for}\ i=1,\dotsc ,n-1\}\ .
\end{equation}
In appendix~\ref{sec:appspectrum} we analyse the structure of the sets $N (d,\epsilon
,k)$. In particular, we show that for any $d$ in the interior of the
fundamental Weyl chamber, there is an $\epsilon_{d}$ such that for
$\epsilon <\epsilon_{d}$ the cardinality $|N
(d,\epsilon ,k)|$ behaves for large $k$ as
\begin{equation}
|N (d,\epsilon ,k)| = \left(\epsilon (k+n+1) \right)^{n-1} +
\mathcal{O} \left((k+n+1)^{n-2} \right)\ .
\end{equation}
The leading term is independent of $d$; therefore the
set $d_{\Lambda +\rho ,\Lambda '+\rho}$ (better: the set of their representatives
in the fundamental chamber) assumes a uniform distribution in the
limit. Although we have chosen here a very specific ``$\epsilon$-box'' of
$d$ to define the average, the results will be independent of the
shape of the neighbourhood of $d$ that is used.

The correlators of the averaged fields $\phi_{d}^{(\epsilon,k)}$ do
not have a well behaved limit. On the other hand, we have the freedom
to change the normalisation of the fields, as well as to rescale the
correlators (corresponding to a rescaling of the vacuum state). Let us
denote the field rescaling by a factor $\alpha$, and the vacuum
rescaling by a factor $\beta$. A bulk two-point function on the sphere
is then rescaled by $\alpha^{2}\beta^{2}$, and a bulk one-point
function on the disk by $\alpha \beta$ (for a more detailed discussion
of these rescalings see~\cite{Runkel:2001ng,Fredenhagen:2007tk}). As
these are the only two correlators we are discussing in this work, we
cannot disentangle the contribution of the different rescalings, and
we will just consider the rescaling of these correlators by the
combination $\gamma =\alpha \beta$.

\noindent The bulk two-point function in the limit theory is then given by
\begin{equation}\label{ansatztwopoint}
\langle \phi_{d_{1}} (z)\phi_{d_{2}} (w)\rangle = \lim_{\epsilon \to 0}
\lim_{k\to \infty} (\gamma (k,n))^{2} \langle \phi_{d_{1}}^{(\epsilon,k)} (z)
\phi_{d_{2}}^{(\epsilon,k)} (w)\rangle \ .
\end{equation}
Here, we chose the normalisation factor $\gamma (k,n)$ to be
independent of $d$. We will now determine $\gamma$ by the requirement
that the bulk two-point function in the limit should be given by
\begin{equation}\label{limittwopoint}
\langle \phi_{d_{1}} (z) \phi_{d_{2}} (w)\rangle 
= \delta (d_{1}-d_{2}^{+}) |z-w|^{-4h_{d_{1}}} \ ,
\end{equation}
where $d_{2}^{+}$ is the label conjugate to $d_{2}$, i.e.\
$(d_{2}^{+})_{i}= (d_{2})_{n-i}$. The conformal weight $h_{d}$ is
obtained as the $k\to \infty$ limit of~\eqref{confweightminmod}, 
\begin{equation}
h_{d} = \frac{1}{2}d^{2} \ .
\end{equation}
When we evaluate~\eqref{ansatztwopoint} by using the
expression~\eqref{minmodtwopoint} for the two-point function in the
minimal models, we obtain
\begin{align}
\langle \phi_{d_{1}} (z)\phi_{d_{2}} (w)\rangle & = \lim_{\epsilon \to 0}
\lim_{k\to \infty} (\gamma (k,n))^{2} (\epsilon (k+n+1))^{-2(n-1)}\nonumber\\
&\quad \times \sum_{(\Lambda_{1};\Lambda_{1}')\in N (d_{1},\epsilon ,k)}
\sum_{(\Lambda_{2};\Lambda_{2}')\in N (d_{2},\epsilon ,k)}
\langle \phi_{(\Lambda_{1};\Lambda_{1}')} (z)\phi_{(\Lambda_{2};\Lambda
_{2}')} (w)\rangle\\
& = \lim_{\epsilon \to 0}
\lim_{k\to \infty} (\gamma (k,n))^{2} (\epsilon (k+n+1))^{-2(n-1)}\nonumber\\
& \quad \times \sum_{(\Lambda_{1};\Lambda_{1}')\in N (d_{1},\epsilon
,k)\cap N (d_{2},\epsilon ,k)^{+}}
|z-w|^{-4h_{\Lambda_{1},\Lambda_{1}'}}\\
& = \lim_{\epsilon \to 0}
\lim_{k\to \infty} (\gamma (k,n))^{2} (k+n+1)^{-(n-1)}\nonumber\\
& \quad \times \prod_{i=1}^{n-1}\left(\epsilon^{-2} (\epsilon
-|d_{1,i}-d_{2,i}|)\Theta (\epsilon -|d_{1,i}-d_{2,i}|) \right)
|z-w|^{-4h_{d_{1}}}\ . 
\end{align}
In the last step we used the result~\eqref{app:intersection} for the intersection of the
two sets $N (d_{i},\epsilon ,k)$. The Heaviside function $\Theta (x)$ is defined
to be $1$ for $x>0$ and $0$ otherwise. The $\epsilon$-dependent term
leads to a delta distribution in the limit,
\begin{equation}\label{deltalimit}
\lim_{\epsilon \to 0}\epsilon^{-2} (\epsilon -|x|)\Theta (\epsilon -|x|) =
\delta (x) \ .
\end{equation}
The coefficients $d_{i}$, $i=1,\dotsc ,n-1$, are the coordinates of $d$
with respect to the fundamental weights $\omega_{i}$, which do not form an
orthonormal basis. The standard inner product on the weight space,
\begin{equation}
(d,d') = \sum_{i=1}^{n-1}d_{i} Q_{ij}d'_{j}\ ,
\end{equation} 
is given by the quadratic form matrix $Q$ with $\det Q = n^{-1}$, so
that the integration measure is
\begin{equation}\label{intmeasure}
d^{n-1}d = \frac{1}{\sqrt{n}}\prod_{i=1}^{n-1} dd_{i}\ .
\end{equation} 
The delta distribution on the weight space is therefore given by
\begin{equation}
\delta (d_{1}-d_{2}) = \sqrt{n}\prod_{i=1}^{n}\delta
(d_{1,i}-d_{2,i})\ .
\end{equation}
If we choose 
\begin{equation}\label{defgamma}
\gamma (k,n) = n^{1/4}(k+n+1)^{(n-1)/2}\ ,
\end{equation}
we obtain the canonically normalised two-point
function~\eqref{limittwopoint} in the limit.

\subsection{Untwisted boundary conditions}
\label{sec:minmodlimuntwist}
Let $s= (s_{1},\dotsc ,s_{n-1})=\sum_{i} s_{i}\omega_{i}$ be a vector
in weight space in the fundamental Weyl chamber, so that the
coefficients $s_{i}$ of the fundamental weights are real non-negative
numbers. We decompose the vector $s$ into its integer $\lfloor
s\rfloor $ part and its fractional part $\{s \}$ (meaning just to take
integer and fractional parts of the coefficients $s_{i}$). Then we
consider the boundary conditions
\begin{equation}\label{scaledbc}
(L_{1};L_{2}) (s,k) = (\lfloor s \rfloor + \lfloor k\{s \}\rfloor
,\lfloor k\{s \}\rfloor  ) \ .
\end{equation} 
Notice that the labels $L_{1}$, $L_{2}$ can lie outside of the
fundamental affine Weyl chambers at level $k$ and $k+1$,
respectively. In that case we reflect the label back to the
fundamental affine Weyl chamber by some affine Weyl transformation,
and consider the corresponding boundary condition. To find the
appropriate bulk one-point function, we observe that the
coefficient~\eqref{mm-one-point} of the one-point function can be
evaluated for arbitrary elements in the fundamental Weyl chamber of
the finite dimensional algebra $sl (n)$. It coincides with the
coefficient for the reflected labels up to a sign, which is determined
by the affine Weyl transformation that is necessary to bring the label
to the fundamental affine chamber. For large level $k$, the necessary Weyl
elements for the numerator label $L$ and the denominator label $L'$
will coincide, so that their signs cancel. Therefore, for large levels
$k$ we can directly use eq.\ \eqref{mm-one-point} for the one-point
functions of the boundary conditions~\eqref{scaledbc}.

When we take the limit $k\to \infty$, we will keep $s$ fixed and scale
the boundary labels $(L;L')$ according to~\eqref{scaledbc}. In the
bulk one-point function, we also keep the bulk label combination
$d=d_{\Lambda+\rho ,\Lambda'+\rho}$ fixed. The exponential term
in the bulk one-point function~\eqref{mm-one-point} then reads
\begin{equation}
e^{-2\pi it^{-1}\big(w (
d_{\Lambda+\rho ,\Lambda'+\rho}),L+\rho \big)+2\pi i\big(w'
(d_{\Lambda+\rho,\Lambda'+\rho}),L'+\rho \big)} = e^{-2\pi i
\big(d,w^{-1} (\lfloor k\{s \}\rfloor ) - w'^{-1} (\lfloor k\{s
\} \rfloor ) \big) + \dotsb } \ .
\end{equation}
In the limit $k\to \infty$ we find strongly oscillating
terms in the bulk one-point functions. On the other hand, a field
$\phi_{d}$ in the limit theory is obtained from the
average~\eqref{defaveragedfields} over bulk fields $\phi_{(\Lambda;\Lambda')}$ with
$d_{\Lambda+\rho,\Lambda'+\rho}\to d$. In averaging the
strongly oscillating terms are suppressed, and we only get
contributions from the terms with $w_{1}=w_{2}$ for which the oscillating
terms cancel (if we assume generic $\{s \}$ -- we shall comment on the
degenerate case below). The prefactor $\mathcal{N}^{1/2}$ that
enters~\eqref{mm-one-point} through~\eqref{defAM} behaves like $k^{- (n-1)/2}$ for large
$k$ (see eq.\ \eqref{defN}). Similarly to our Ansatz for the limit of the bulk two-point
function in~\eqref{ansatztwopoint}, we have to rescale the one-point function by the
factor $\gamma (k,n)$ (see~\eqref{defgamma}) to obtain the one-point
function in the limit theory,
\begin{align}
\tilde{U}^{M}_{s} (d) & := \lim_{k\to \infty} \gamma (k,n)
A_{M} (d)
\sum_{w \in W} e^{-2\pi i \big(w (d),\lfloor s\rfloor +
(t^{-1}-1)\lfloor k\{s \}\rfloor    \big)} \\
&=\prod_{e>0}\left| 2\sin \left(\pi (e,d)
\right)\right|^{-1}
\sum_{w \in W} e^{-2\pi i \big(w (s),d  \big)}\ .
\label{minmodgenericonepoint}
\end{align}
Note that the so obtained one-point coefficient is the same on the
whole Weyl orbit of $d$, so it is independent of which representative
we choose. When we compare to the one-point
functions~\eqref{limitnondeg} that we obtained from the non-degenerate
boundary conditions in Toda theory, we find coincident results if we
identify the bulk fields $V_{ip}$ from Toda theory and $\phi_{d}$ from
the minimal models by
\begin{equation}
\phi_{d} \leftrightarrow \pm i^{n (n-1)/2} (\pi
\mu_{\text{ren}})^{(d,\rho)} V_{id} \ .
\end{equation}
This identification is also consistent with the two-point
functions~\eqref{Todatwopointlimit} and~\eqref{limittwopoint}. 

The above result~\eqref{minmodgenericonepoint} was derived for generic
$s$. If $\{s \}$ sits on a boundary of the fundamental Weyl chamber,
there are some Weyl reflections that leave it invariant, in other
words, the stabiliser group $W_{\{s \}}$,
\begin{equation}
W_{\{s \}}= \big\{ w\in W \, |\,  w (\{s \}) = \{s \} \big\} \ ,
\end{equation}
is non-trivial. In that case, requiring the strongly oscillating terms
to cancel leads to the condition $w_{2}w_{1}^{-1}\in W_{\{s \}}$ (instead
of $w_{1}=w_{2}$ for generic $s$). The limit then becomes
\begin{equation}
\tilde{U}^{M}_{s} (d) = \prod_{e>0}\left|2\sin \left(\pi (e,d)
\right)\right|^{-1}  \sum_{w\in W,w'\in W_{\{ s\}}}
\epsilon (w') e^{-2\pi i\big(w(s+\rho -w' (\rho)),d\big)}\ .
\end{equation}
This reproduces the degenerate one-point functions
$\tilde{U}^{T,W'}_{\kappa ,\Omega ,\Omega '}$ of Toda theory (see eq.\
\eqref{limitdeg}) with $W'=W_{\{s \}}$, $\kappa +\Omega =s$ and $\Omega
'=0$. 

In the completely degenerate case ($\{s \}=0$), the boundary labels 
$(L;L') = (s;0)$ are kept fixed in the limit. One might expect
that one could get more boundary states by choosing arbitrary
$(L;L')$ with non-trivial $L'$ and keeping the labels fixed
in the limit. These, however, do not lead to new boundary conditions,
as we shall see now.

\noindent Keeping the labels fixed, we obtain the one-point function
\begin{align}
\tilde{U}^{M}_{(L;L')} (d) & = \prod_{e>0}\left|2\sin \left(\pi (e,d)
\right)\right|^{-1}
\sum_{w\in W} \epsilon (w) e^{-2\pi i\big(w (
d),L+\rho \big)} \nonumber\\
&\quad \times \sum_{w' \in W} \epsilon (w') e^{2\pi i \big(w'
(d),L'+\rho \big)} \\
& =  \prod_{e>0}\left|2\sin \left(\pi (e,d)
\right)\right|\chi_{L} (-2\pi i d)\chi_{L'} (2\pi i d)\ ,
\end{align}
where $\chi_{L}$ is the finite $sl (n)$ character of the
representation with highest weight $L$. The product of the characters
is just given by the tensor product rules,
\begin{equation}
\chi_{L} (-2\pi i d)\chi_{L'} (2\pi i d) = \sum_{L''}
N_{LL'^{+}}{}^{L''} \chi_{L''} (-2\pi i d) \ ,
\end{equation}
where $L'^{+}$ labels the representation conjugate to $L'$. For
the one-point function coefficients this means
\begin{equation}
\tilde{U}^{M}_{(L;L')} (d) = \sum_{L''} N_{LL'^{+}}{}^{L''}
\tilde{U}^{M}_{(L'';0)} (d)\ ,
\end{equation}
i.e.\ the boundary condition $(L;L')$ can be decomposed into a superposition
of boundary conditions of the form $(L'';0)$ in the limit.\footnote{\label{footnoteflows}It
is known that in a minimal model the boundary state labelled by
$(L;L')$ can flow to a superposition of $(L'';0)$ boundary states, and
that this boundary renormalisation group flow can be described in
perturbation theory in $1/k$ for large levels, becoming shorter and
shorter for higher levels~\cite{Fredenhagen:2001kw}
(see~\cite{Recknagel:2000ri,Graham:2001tg} for the Virasoro
case). Also from this perspective it can be expected that these
boundary configurations are identified in the limit.}
\smallskip

We have seen that we can reproduce both the generic and the various
degenerate boundary conditions that we obtained from Toda theory. The
only mismatch seems to be that we do not get the generic Toda boundary
condition $\tilde{U}^{T}_{s}$ for a vector $s$ whose fractional part
is degenerate (and similar situations for partially degenerate
boundary conditions). Is there something special about those boundary
conditions? Let us take $s=\Omega$ as an integral weight, so its
fractional part is completely degenerate. We then claim that the
non-degenerate boundary condition $\tilde{U}^{T}_{s}$ (given
in~\eqref{limitnondeg}) for this value
of $s$ decomposes into an infinite collection of completely degenerate
boundary conditions~\eqref{limitcompdeg},
\begin{equation}\label{decomposition}
\tilde{U}^{T}_{s=\Omega} (p) = \sum_{\{m_{e} \}\in \mathbb{N}_{0}^{n
(n-1)/2}} \tilde{U}^{T}_{\Omega +\sum_{e>0}m_{e}e,0} (p) \ .
\end{equation}
We rewrite this equation by inserting the
expressions~\eqref{limitnondeg} and~\eqref{limitcompdeg},
\begin{equation}
\sum_{w\in W} e^{-2\pi i (\Omega ,w (p))} = \sum_{\{m_{e} \}\in \mathbb{N}_{0}^{n
(n-1)/2}}\sum_{w'\in W} \epsilon (w')
\sum_{w\in W} e^{-2\pi i (\Omega +\sum_{e>0}m_{e}e+\rho -w' (\rho),w (p))} \ .
\end{equation}
This equality then follows from
\begin{align}
\sum_{\{m_{e} \}\in \mathbb{N}_{0}^{n (n-1)/2}} e^{-2\pi i (\sum_{e>0}m_{e}e,p)} & = 
\prod_{e>0} \left( \sum_{m_{e}\geq 0} e^{-2\pi i (m_{e}e,p)}\right) \\
& = \prod_{e>0} e^{\pi i (e,p)}\left(e^{\pi i (e,p)}-e^{-\pi i (e,p)} \right)^{-1} \\
& = e^{2\pi i (\rho ,p)} \left(\sum_{w'\in W}\epsilon (w') e^{2\pi i
(w' (\rho),p)} \right)^{-1} \ .
\end{align}
Here we used the Weyl denominator formula (see e.g.\
\cite{FrancescoCFT}).

When the fractional part of $s$ degenerates, the generic boundary
condition turns into a superposition of degenerate ones. This
phenomenon is known~\cite{Fredenhagen:2004cj} from the $sl (2)$ case,
where in the limit $c\to 1$, the (generic) FZZT boundary condition of Liouville theory
decomposes into an infinite array of (degenerate) ZZ boundary
conditions for a discrete set of boundary parameters.

\subsection{Twisted boundary conditions}
 
We have seen in section~\ref{sec:minmodtwisted} that twisted boundary conditions in the
$sl(2m+1)$ cosets can be labelled by self-conjugate weights of
$sl(2m+1)$. To describe twisted boundary conditions in the limit $k\to
\infty$, we now choose a self-conjugate vector $s=s^{+}$ in the
fundamental Weyl chamber of the weight space of $sl (2m+1)$. With this
we associate the boundary condition
\begin{equation}\label{scaledtwistedbc}
(L;L') = (\lfloor s\rfloor +\lfloor k\{s \}\rfloor;\lfloor k\{ s \}\rfloor )\ .
\end{equation}
The weights $L$ and $L'$ are not necessarily in the fundamental affine
Weyl chamber of $sl(2m+1)$ at the levels $k$ and $k+1$, respectively.
They can be reflected back by the use of the affine Weyl group, or
better by the subgroup that maps self-conjugate labels to
self-conjugate ones (for details see the discussion at the end of
appendix~\ref{sec:apptwistedS}). The
coefficients~\eqref{twistedbc} of the boundary state are invariant under these
reflections up to signs. For large enough level $k$, the Weyl
reflections used for $L$ and $L'$ will coincide so that the signs
cancel.

We then define the one-point functions for twisted boundary conditions
in the limit theory as the limit of the one-point
functions~\eqref{twistedbc} of
averaged fields~\eqref{defaveragedfields} with boundary
conditions~\eqref{scaledtwistedbc} where we take $s$ fixed. We obtain
\begin{align}
\tilde{U}^{M}_{\omega ,s} (d) & = \lim_{\epsilon \to 0}\lim_{k\to \infty}
\gamma (k,n) \frac{1}{|N (d,\epsilon ,k)|} 
\sum_{(\Lambda ;\Lambda')\in N (d,\epsilon ,k)} 
U^{M}_{\omega ,(\lfloor s\rfloor +\lfloor k\{s \}\rfloor,\lfloor k\{ s \}\rfloor
)} (\Lambda ;\Lambda ')\\
& = n^{1/2}\prod_{e>0}\left|2 \sin (\pi (e,d))\right|^{-1}
\lim_{\epsilon \to 0}\lim_{k\to \infty}k^{m}
 \frac{1}{|N (d,\epsilon ,k)|}  
\sum_{(\Lambda ;\Lambda')\in N (d,\epsilon ,k)} \delta_{(\Lambda
;\Lambda ') (\Lambda^{+};\Lambda '^{+})}\nonumber\\
& \quad \times 
\sum_{w,w'\in W^{\omega}}
e^{-\pi i\left(t^{-1}w (\lfloor s\rfloor +\lfloor k\{s\}\rfloor+\rho)
-w' (\lfloor k \{ s\} \rfloor+\rho ),d_{\Lambda +\rho ,\Lambda '+\rho}\right)}
\ ,
\end{align}
where the normalisation factor $\gamma$ was given in~\eqref{defgamma}. 
Similarly to the arguments for the limit of untwisted boundary
conditions, we observe a strongly oscillating behaviour if $w\not=
w'$ and generic $s$. The averaging over $d$ suppresses these terms, so
that in the limit we are left with the contributions from $w=w'$. On
the other hand, the condition that the field labels $(\Lambda;\Lambda
')$ are self-conjugate restricts the sum over $N (d,\epsilon ,k)$ to a subset of size
\begin{equation}
\left| \{(\Lambda ;\Lambda ')\in N (d,\epsilon ,k)\, |\, (\Lambda
;\Lambda ')= (\Lambda^{+};\Lambda '^{+})\} \right| = k^{m}\prod_{j=1}^{m}
\left(\epsilon  -|d_{i}-d_{n-i}| \right)\Theta (\epsilon
-|d_{i}-d_{n-i}|) + \dotsb 
\end{equation}
where we left out subleading contributions in $k$ (see eq.\
\eqref{app:Nomega}).  Upon sending $\epsilon \to 0$, we obtain a
product of delta distributions (see~\eqref{deltalimit}), and the
one-point coefficient is given by
\begin{equation}\label{gentwistedlimitbcpre}
\tilde{U}^{M}_{\omega ,s} (d) = n^{1/2}\prod_{e>0}\left|2 \sin (\pi
(e,d))\right|^{-1} \prod_{i=1}^{m}\delta (d_{i}-d_{n-i})
\sum_{w \in W^{\omega}}
e^{-\pi i \left(w (s),d \right)} \ .
\end{equation}
We would like to rewrite the delta distributions. The weight space $V$
can be decomposed as an orthogonal sum of self-conjugate and anti
self-conjugate vectors,
\begin{equation}
V=V_{S} \oplus V_{A}\ .
\end{equation}
The measure $d^{2m}v= (d^{m}v_{S}) (d^{m}v_{A})$ factorises, and on the
symmetric vectors parameterised by $(v_{1},\dotsc ,v_{m},v_{m},\dotsc ,v_{1})$ the
measure takes the form
\begin{equation}
d^{m}v_{S} = 2^{m/2}\prod_{j=1}^{m} dv_{j} \ ,
\end{equation}
because the quadratic form matrix $\tilde{Q}$, $(v_{S},v'_{S})=
\sum_{i,j=1}^{m} v_{i}\tilde{Q}_{ij}v_{j}$, has determinant $2^{m}$ (it is
related to the quadratic form matrix $\hat{Q}$ of $sp (2m)$ by
$\tilde{Q}_{ij}=4\hat{Q}_{ij}$, and $\det \hat{Q}=2^{-m}$). The
measure on the full space is given by~\eqref{intmeasure}. We then
have 
\begin{align}
\int d^{2m}v \left( n^{1/2} \prod_{j=1}^{m} \delta (v_{i}-v_{n-i})\right)f (v)
&=\int \prod_{i=1}^{2m}dv_{i} \prod_{j=1}^{m}\delta (v_{i}-v_{n-i})f (v) \\
& = \int \prod_{i=1}^{m} dv_{i} f (v_{S}) \\
& = \int d^{m}v_{S}\, 2^{-m/2}f (v_{S}) 
\end{align}
for an arbitrary function $f$, so that
\begin{equation}
2^{-m/2}\delta^{(m)} (v_{A}) = n^{1/2} \prod_{j=1}^{m} \delta (v_{i}-v_{n-i})\ .
\end{equation}
The one-point coefficient~\eqref{gentwistedlimitbcpre} hence becomes
\begin{equation}\label{gentwistedlimitbc}
\tilde{U}^{M}_{\omega ,s} (d) = \delta^{(m)} (d_{A}) 2^{-m/2} 
\prod_{e>0}\left|2 \sin (\pi(e,d))\right|^{-1} 
\sum_{w \in W^{\omega}} e^{-\pi i \left(w (s),d \right)} \ .
\end{equation}
This is the answer for generic $s$. If the stabiliser subgroup
$W_{s}^{\omega}\subset W^{\omega}$,
\begin{equation}
W_{s}^{\omega} = \{w\in W^{\omega}\, |\, w (\{s \})=\{s \}  \}\ ,
\end{equation}
is non-trivial, we obtain degenerate boundary conditions with
one-point functions determined by
\begin{equation}\label{degtwistedlimitbc} 
\tilde{U}^{M}_{\omega ,s} (d) = \delta^{(m)} (d_{A}) 2^{-m/2}
\prod_{e>0}\left|2 \sin (\pi (e,d))\right|^{-1}
\sum_{w \in W^{\omega},w'\in W^{\omega}_{s} } \epsilon (w')
e^{-\pi i \left(w (s+\rho -w' (\rho)),d \right)} \ .
\end{equation}
In the completely degenerate case ($\{s \}=0$), the boundary labels
$(L;L')= (s;0)$ are kept fixed in the limit. As in the case of
untwisted boundary conditions, one might ask the question whether one
obtains further boundary conditions by taking $(L;L')$ fixed in the limit
with a non-trivial label $L'$. These lead to a one-point
coefficient
\begin{align}
\tilde{U}^{M}_{\omega, (L;L')} &= \delta^{(m)} (d_{A}) 2^{-m/2} \prod_{e>0}\left|2 \sin (\pi
(e,d))\right|^{-1} \sum_{w\in W^{\omega }} \epsilon (w)
e^{-\pi i\left(w (d),L+\rho \right)}\nonumber\\
&\quad \times \sum_{w'\in W^{\omega }} \epsilon (w')
e^{\pi i \left(w' (d),L'+\rho \right)} \ .
\label{compdegtwistedlimitbc}
\end{align}
The symmetric subgroup $W^{\omega}\subset W$ is isomorphic to the Weyl
group $\hat{W}$ of the finite dimensional algebra $sp (2m)$. The sums
in~\eqref{compdegtwistedlimitbc} can therefore rewritten in terms of characters
of $sp (2m)$, 
\begin{align}
\tilde{U}^{M}_{\omega, (L;L')} & = \delta^{(m)} (d_{A}) 2^{-m/2} \prod_{e>0}\left|2 \sin (\pi
(e,d))\right|^{-1} \prod_{\hat{e}>0} 
\left(2\sin \left( 2\pi \left( \hat{e},\hat{d}
\right)\right) \right)^{2}\nonumber\\
& \quad \times \chi_{\hat{L}} \left( -4\pi i \hat{d}\right)
\chi_{\hat{L}'} \left( 4\pi i\hat{d} \right)\ ,
\end{align}
where $\hat{L}$, $\hat{L}'$ and $\hat{d}$ are vectors in the weight
space of $sp (2m)$ such that $\hat{L}_{i}=L_{i}$ for $i=1,\dotsc ,m$
and so on. By $\hat{e}$ we denote the roots of $sp (2m)$.
The product of characters can be decomposed into a sum of characters
using the tensor product rules\footnote{The appearance of the $sp (2m)$ tensor product
rules might be expected from the analysis of twisted D-brane charges
in $SU (2m+1)$ WZW models~\cite{Alekseev:2002rj,Gaberdiel:2003kv}.} of $sp (2m)$,
\begin{equation}
\chi_{\hat{L}} \left(-4\pi i \hat{d}\right) \chi_{\hat{L}'} \left(4\pi
i \hat{d}\right) 
= \sum_{\hat{L}''}N_{\hat{L} \hat{L}'}{}^{\hat{L}''}
\chi_{\hat{L}''} \left(-4\pi i\hat{d}\right) \ .
\end{equation}
Notice that the $sp (2m)$ representations are self-conjugate, so that
$\chi_{\hat{L}} (\xi)= \chi_{\hat{L}} (-\xi)$. We conclude that in the
limit theory the twisted boundary condition labelled by $(L;L')$ can
be identified\footnote{Similarly to the discussion in
footnote~\ref{footnoteflows} on page~\pageref{footnoteflows}, this
identification is expected from the work
of~\cite{Fredenhagen:2001kw,Alekseev:2002rj,Fredenhagen:thesis}, from
which one can show that a twisted boundary state in a minimal model
labelled by $(L;L')$ flows to a superposition of boundary states
$(L'';0)$, and that this flow is perturbative in $1/k$.} with a
superposition of boundary conditions $(L'';0)$,
\begin{equation}
\tilde{U}^{M}_{\omega ,(L;L')} (d) = \sum_{L''=L''^{+}}
N_{\hat{L}\hat{L}'}{}^{\hat{L}''}\tilde{U}^{M}_{\omega ,(L'';0)} (d) \ . 
\end{equation}

\section{Conclusion}

We have analysed untwisted boundary conditions in $sl (n)$ Toda CFTs
and in $W_{n}$ minimal models in the limit $c\to n-1$. The expressions
for the one-point function in the presence of these boundary
conditions agree. Furthermore, we have studied the limit of twisted
boundary conditions in $W_{n}$ minimal models for odd $n=2m+1$. The
results~\eqref{gentwistedlimitbc} and~\eqref{degtwistedlimitbc} for
the twisted one-point functions in the limit theory suggest a
generalisation to Toda theory. An obvious guess for the non-degenerate
twisted one-point coefficients in Toda theory would be
\begin{equation}\label{proposal}
U^{T}_{\omega ,s} (\alpha ) = \delta^{(m)} (p_{A}) 2^{-m/2}
A (\alpha)^{-1} \sum_{w\in
W^{\omega}} e^{-\pi \left(w (s),\alpha -Q \right)} \ . 
\end{equation}
Here, the boundary parameter $s$ is a symmetric weight vector, $\alpha
=Q+ip$, and $p_{A}$ is the anti-symmetric part of $p$ under
conjugation ($p$ is considered to be in the fundamental Weyl chamber).
Similarly one is led to proposals for the degenerate boundary
conditions. In particular, the completely degenerate twisted boundary
condition is expected to be given by
\begin{equation}
U^{T}_{\omega , (L,L')} (\alpha) = \delta^{(m)} (p_{A}) 2^{-m/2} 
A (\alpha)^{-1}
\sum_{w,w' \in W^{\omega}}\epsilon (w')  
e^{\pi i \left(w \left(b (L+\rho)+b^{-1}w' (L'+\rho) \right),
\alpha -Q \right)} \ , 
\end{equation}
where $L$ and $L'$ are self-conjugate, dominant, integral weights of $sl (n)$.
It is not hard to see that these proposals are consistent with the
analysis of the light asymptotic limit of such one-point functions
in~\cite{Fateev:2010za}.

We have not discussed the boundary spectrum in this work. In the
Virasoro case ($n=2$), it turns out that the boundary spectrum has an
interesting band structure~\cite{Fredenhagen:2004cj} that varies with
the parameter $s$. A similar story is expected for higher $n$. We saw
a glimpse of it at the end of section~\ref{sec:minmodlimuntwist} when
we observed the decomposition~\eqref{decomposition} of generic
boundary conditions into infinite collections of degenerate ones when
the fractional part $\{s \}$ of the boundary label degenerates. For
$\{s \}=0$ the boundary spectrum therefore becomes discrete, and for
general parameters $s$ one expects that the spectrum is contained in
some bands which degenerate for integral values of $s$. The complete
analysis will be more complicated than in the $sl (2)$ case, because
the fusion rules are more complicated, and in particular non-trivial
multiplicities appear that might diverge in the limit. Still, this
analysis could be helpful to understand the divergences of the annulus
partition functions in~\cite{Fateev:2010za} also from the minimal
model side.  
\smallskip

There are several ways to generalise and extend our analysis. First of
all, the results should have a straightforward generalisation to
minimal models and Toda theories based on other simply-laced Lie
algebras.  Another extension would be to study defects in minimal
models and Toda theories and in their common limit. All maximally
symmetric, topological defects in the minimal models can be obtained
by the standard constructions, and their limit can be taken following
the steps that we used for the boundary conditions. This limit can be
compared to the Toda theories. There, untwisted maximally symmetric
defects in Toda theories have been described
in~\cite{Drukker:2010jp}. The twisted ones should have a very similar
form, and should also resemble the twisted boundary
conditions~\eqref{proposal}. Indeed, for $sl (2m+1)$ the above form~\eqref{proposal}
suggests the generic twisted topological defect to be
\begin{align}
\mathcal{O}_{s} &= \int dp_{S} 2^{-m/2}\frac{\sum_{w\in W^{\omega}}e^{-\pi (w
(s),\alpha -Q)}}{\prod_{e>0}\left(-4\sin \pi b (\alpha-Q ,e) \sin \pi
b^{-1} (\alpha-Q ,e) \right)} \\
&\quad \times \sum_{\{k \},\{l \}} |\alpha ,k;\alpha ,l\rangle \otimes 
\langle \alpha ,k;\alpha ,l^{+}| \ ,
\end{align}
where the integral goes over self-conjugate labels $p=i (Q-\alpha)$ in
the fundamental Weyl chamber. The states $|\alpha ,k\rangle$ form an
orthonormal basis in the $W_{n}$ representation based on a ground
state $|\alpha \rangle$, so that $\sum_{k}|\alpha ,k\rangle \langle
\alpha ,k|$ is a projector on this representation. Similarly
$\sum_{l}|\alpha ,l\rangle \langle \alpha ,l^{+}|$ is a ``twisted''
projector that implements the charge conjugation twist. This defect
would then correspond to having trivial gluing conditions for the
holomorphic currents and twisted gluing conditions for the
anti-holomorphic ones. These twisted defects could also be of interest
for the relation between Toda theories and four-dimensional
supersymmetric gauge theories~\cite{Drukker:2010jp,Tachikawa:2010vg}.

Another way of extending the analysis of the limit theory would be to consider bulk
three-point correlators. In Toda theories they are only known for a
subset of primary fields~\cite{Fateev:2007ab,Fateev:2008bm}. For
minimal models, the structure constants can be obtained from a free
field construction~\cite{Lukyanov:1990tf}, but the expressions contain
integrals over the screening charges. It is not known how to
explicitly evaluate these integrals for arbitrary fields, but if one
of the field labels takes a special form, the integrals can be
evaluated following the work of~\cite{Fateev:2007ab}. It
would be interesting to compare at least these accessible data in Toda
theory and minimal models. 
\smallskip

Recently, limits of $W_{n}$ minimal models have been investigated in
the context of AdS/CFT duality~\cite{Gaberdiel:2010pz} (see
also~\cite{Kiritsis:2010xc}).  There, in
addition to sending the level $k$ to infinity, also the rank of the
algebra grows, while the ratio $\lambda =\frac{n}{k+n}$ is kept fixed
($\lambda$ takes the role of a `t~Hooft coupling). Sending first $k\to
\infty $ while keeping $n$ fixed, as we did in our analysis, and then
sending $n\to \infty$ would correspond to zero `t Hooft coupling
$\lambda=0$. However, the analysis of~\cite{Gaberdiel:2010pz} shows
that the $\lambda \to 0$ limit of their theories corresponds to a
limit theory that is different from ours, in particular it should have
discrete spectrum. The relevant limiting procedure for the $\lambda=0$
case of~\cite{Gaberdiel:2010pz} therefore seems to be the $k\to
\infty$ limit in the sense of~\cite{Roggenkamp:2003qp}, followed by
the $n\to \infty$ limit.

\section*{Acknowledgements}

I would like to thank Matthias Gaberdiel and Volker Schomerus for
useful comments on the draft of this paper.

\begin{appendix}
\section{The spectrum of minimal models in the limit}
\label{sec:appspectrum}
In section~\ref{sec:minmodlimitbulk} we introduced the set $N(d,\epsilon ,k)$ 
(see eq.\ \eqref{defofN}) of minimal model labels $(\Lambda ;\Lambda ')$ whose
associated Weyl orbit of the weight vector $d_{\Lambda +\rho,\Lambda
'+\rho}$ has a representative $w (d_{\Lambda +\rho ,\Lambda '+\rho})$ close
to $d$ in the sense that it is contained in an ``$\epsilon$-box'' around $d$,
\begin{equation}\label{app:closetod}
\left| (w(d_{\Lambda +\rho ,\Lambda '+\rho}))_{i} 
- (d)_{i}\right| <\frac{\epsilon}{2} \quad \text{for}\ j=1,\dotsc ,n-1\ .
\end{equation}
We here want to analyse the structure of these sets, and in particular
determine their cardinalities and the cardinalities of their
intersections.

\begin{figure}
\begin{center}
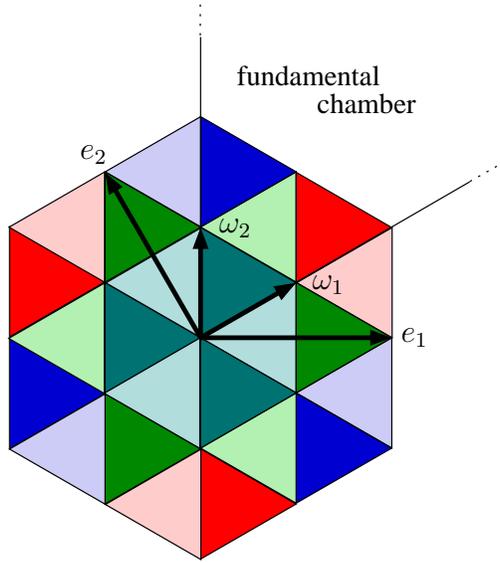
\end{center}
\caption{\label{fig1}This is the weight space of $sl (3)$ with
fundamental weights $\omega_{1}$, $\omega_{2}$ and simple roots
$e_{1}$, $e_{2}$. The dark regions contain those vectors $d$
satisfying $\sum_{i=1}^{n-1}\{d_{i} \}\leq 1$. The Weyl orbit of one
triangular region contains three dark triangles with vectors satisying
the condition, and three light triangles. For any vector $d$ there are
therefore Weyl images $w (d)$ satisfying the condition, and these
images are related by the group $I_{W}$ given in~\eqref{app:defofI}, which is the
$\mathbb{Z}_{3}$ rotation group in the case of $sl (3)$.}
\end{figure}

\noindent First let us look at the structure of $d_{\Lambda +\rho ,\Lambda
'+\rho}$. We have
\begin{equation}
d_{\Lambda +\rho ,\Lambda '+\rho} = (\Lambda -\Lambda ') +
\frac{1}{k+n+1} (\Lambda '+\rho) \ ,
\end{equation}
so that the integer and fractional parts of $d_{\Lambda +\rho ,\Lambda
'+\rho}$ are given by
\begin{equation}
\lfloor d_{\Lambda +\rho ,\Lambda'+\rho} \rfloor
= \Lambda -\Lambda ' \qquad ,\qquad 
 \{d_{\Lambda +\rho,\Lambda '+\rho} \} = 
\frac{1}{k+n+1} (\Lambda '+\rho) \ .
\end{equation}
We observe that for an integral dominant weight $\Lambda '$ of $sl(n)$
at level $k+1$, we have the restriction
\begin{equation}
\sum_{i=1}^{n-1} \{d_{\Lambda +\rho,\Lambda '+\rho} \}_{i} \leq  1 \ .
\end{equation}
A generic weight vector $d$ does not satisfy this condition, but for
any vector $d$ there is a Weyl transformation $w \in W$ such that 
$w(d)$ is in accord with the restriction,
\begin{equation}
\sum_{i=1}^{n-1} \{w (d) \}_{i} \leq   1 \ .
\end{equation}
To see this, think of the action of the affine Weyl group at level
$1$. Its action on the weights is generated by
translations by vectors in the (co-)root lattice, which leave the
fractional part invariant, and by ordinary Weyl
transformations. For any vector $d$ there is an affine Weyl
transformation sending $d$ to a vector $d'$ in the fundamental affine chamber
$C_{0}$ at level $1$ such that $d'_{i}\geq 0$ and $\sum d'_{i}\leq 1$. 
We have illustrated this for the case of $sl (3)$ in
figure~\ref{fig1}.

Assume now that $d'=w (d)$ is such that $\sum_{i}\{d' \}_{i}\leq  1$. We
further assume that $\{d' \}$ does not sit at the boundary of the
affine Weyl chamber $C_{0}$, i.e.\ that $\{d' \}_{i}>0$ and
$\sum_{i}\{d' \}_{i}<1$; we will comment on the general case later.
The labels $\Lambda ,\Lambda '$ with 
$d_{\Lambda +\rho ,\Lambda '+\rho}$ close to $d'$ in the sense
of~\eqref{app:closetod} then satisfy
\begin{equation}\label{app:inequforLp}
\Lambda = \Lambda ' + \lfloor d' \rfloor \quad ,\quad 
\{d' \}_{i}-\frac{\epsilon}{2} < \frac{\Lambda '_{i}}{k+n+1} <
\{d'\}_{i}+\frac{\epsilon}{2} \ .
\end{equation}
Here we want to assume that $\epsilon$ is small enough so that
the $\epsilon$-box around $\{d' \}$ is still contained in the
interior of $C_{0}$. 
Then the labels $\Lambda '$ that are allowed by the inequality
in~\eqref{app:inequforLp} are dominant weights at level $k+1$. Furthermore we
assume that $k$ is large enough so that $\Lambda$ is a dominant
weight at level $k$. 

\noindent The number of labels $\Lambda '$ that satisfy~\eqref{app:inequforLp}
is then given by
\begin{equation}\label{app:sizeofN}
|N (d,\epsilon ,k) | = \left(\epsilon (k+n+1) \right)^{n-1} +
\mathcal{O} (k^{n-2})\ .
\end{equation}
This describes the behaviour of the cardinality at small but fixed $\epsilon$
and $k$ going to infinity. Note that the $\epsilon$-box around $d'$
that we used to derive~\eqref{app:inequforLp} is
not precisely the Weyl image of the $\epsilon$-box around $d$, but for
the counting only the volume of the box matters.

The result~\eqref{app:sizeofN} was derived for a generic $d$ such that
$\{w (d) \}$ is contained in the interior of the fundamental affine Weyl
chamber $C_{0}$ at level $1$ for some $w \in W$. There are two issues where we
should have been more careful. In the derivation we only considered
one Weyl image $w (d)$, and we have to make sure that we do not get
further contributions from other images. On the other hand, we have
not considered the identification rules~\eqref{fieldidentification}
for the coset labels, which might lead to an overcounting.

Let us therefore analyse which $w'\in W$ map a vector $d'$ with $\{d'\}$ 
inside the fundamental chamber $C_{0}$ at level $1$ to a vector $w' (d')$
with $\{w' (d')\}$ also lying in $C_{0}$. We denote the group
consisting of such $w'$ by $I_{W}$. Given a generic
vector $d$ there is always a unique affine Weyl transformation
consisting of a Weyl rotation $w$ and a translation by an
element $\alpha$ of the (co-)root lattice that maps it to the
fundamental affine 
chamber $C_{0}$,
\begin{equation}\label{app:firstw}
w (d) + \alpha \in C_{0} \ \Rightarrow\ \{w (d) \}  \in C_{0} \ .
\end{equation}
The affine Weyl transformations only include translations by vectors
of the root lattice. The fractional part $\{d \}$ is also invariant under
translations by vectors from the weight lattice. To find another 
$w'\in W$ such that $\{ w' (d)\}\in C_{0}$ we add any integral weight
vector $v$ to $d$, and then look for the unique affine Weyl
transformation $(w',\alpha ')$ mapping it to the fundamental chamber,
\begin{equation}\label{app:secondw}
w' (d+v) +\alpha ' \in C_{0} \ \Rightarrow \ \{w' (d) \} \in C_{0}\ .
\end{equation}
In that way we have found another $w'$ that does the job. The number
of such Weyl group elements is given by the number of different
integral weight
vectors $v\in L_{W}$ modulo the root lattice $L_{R}$. For $sl (n)$
this number is given by $|L_{R}/L_{W}|=n$.  The group $I_{W}$ then
consists of those Weyl elements that map the fundamental affine chamber
$C_{0}$ back to itself modulo the weight
lattice $L_{W}$. From~\eqref{simplecurrentaction} we see that the action of
the simple currents at level $1$ precisely has this
property. Therefore the group $I_{W}$ is generated by 
the Weyl group element $w_{J}$ given in~\eqref{simplecurrentWeyl},
\begin{equation}\label{app:defofI}
I_{W} = \{w_{J}^{j}\, | \, j=0,\dotsc ,n-1 \} \ .
\end{equation}
For any vector $d$ we thus find $n$ different Weyl transformations
$w_{i}$ with $\{w_{i} (d)\}\in C_{0}$, and they are related by the
action of the group $I_{W}$ (this is illustrated in
figure~\ref{fig1}). On the other hand, transformations of the vector
$d_{\Lambda +\rho ,\Lambda '+\rho}$ by a Weyl element $w_{J}^{j}\in
I_{W}$ corresponds to considering the field identification of the
label pair $(\Lambda ;\Lambda ')$ by the simple current $J^{j}$ (see
eq.\ \eqref{fieldidentreltoWeyl}). The additional contributions due to
the appearance of $I_{W}$ are precisely cancelled by dividing out the
field identifications, and the formula~\eqref{app:sizeofN} is
therefore correct.  \smallskip

What does happen if $d$ is non-generic such that $\{d \}$ sits at the
boundary of $C_{0}$? Let us assume that $d$ itself is in the interior
of the fundamental Weyl chamber, and also its $\epsilon$-box. If now
$\{d \}$ (or one of its Weyl images) is on the boundary of $C_{0}$, this
just means that for different parts of the $\epsilon$-box around $d$
we must use different Weyl transformations to bring their fractional
parts to $C_{0}$. The overall counting does not change.

For vectors $d$ that sit at the boundary of the fundamental chamber,
the situation is different because for any arbitrarily small $\epsilon$,  
different parts of the $\epsilon$-box around $d$ have to be identified
and the counting changes. We do not investigate here what
this implies for the spectrum of the limit theory.
\smallskip

\noindent Let us now discuss the intersection of two sets, 
\begin{equation}
N (d_{1},\epsilon,k) \cap N (d_{2},\epsilon ,k) \ ,
\end{equation}
where $d_{1}$ and $d_{2}$ are in the interior of the fundamental Weyl
chamber, and $\epsilon$ is small enough such that the counting we have done
above works both for $d_{1}$ and $d_{2}$. 

Assume that $d_{1}$ is such that $\{d_{1} \}$ is in the interior of
$C_{0}$. For $d_{2}$ we consider the vectors $w (d_{2})$ with $\{w
(d_{2}) \}\in C_{0}$, and all the label pairs $\Lambda ,\Lambda '$
such that $d_{\Lambda +\rho ,\Lambda '+\rho}$ is in the
$\epsilon$-box around $w (d_{2})$. Now the question is: which of them
coincide with labels $\Lambda ,\Lambda '$ with $d_{\Lambda
+\rho ,\Lambda '+\rho}$ being contained in the $\epsilon$-box around $d_{1}$? To have
coincident labels $\Lambda '$ we need that $\{d_{1} \}$ and $\{w
(d_{2}) \}$ are close, i.e.\ their $\epsilon$-boxes must have a
non-empty intersection. To also have coincident labels $\Lambda$, we
need that $\lfloor d_{1}\rfloor $ and $\lfloor w(d_{2})\rfloor$
coincide. Therefore the $\epsilon$-boxes around $d_{1}$ and 
$w(d_{2})$ have to overlap, which can only occur for $w=1$.

\noindent The counting of coincident labels in the intersection then results in 
\begin{equation}
\label{app:intersection}
|N (d_{1},\epsilon ,k)\cap N (d_{2},\epsilon ,k)| = \prod_{i=1}^{n-1}
\left((k+n+1) 
(\epsilon - | (d_{1}-d_{2})_{i}|)\Theta (\epsilon - |
(d_{1}-d_{2})_{i}|)  \right) + \dotsb \ .
\end{equation}
Here, the Heaviside function $\Theta$ is defined such that $\Theta
(x)=1$ for $x>0$ and $\Theta (x)=0$ otherwise, and it encodes the
condition that the two $\epsilon$-boxes intersect.
\smallskip

When we discuss twisted boundary conditions, we also have to analyse
how many self-conjugate labels are contained in $N (d,\epsilon ,k)$,
i.e.\ the cardinality of the set
\begin{equation}
N^{\omega} (d,\epsilon ,k) = \{ (\Lambda ;\Lambda ')\in N (d,\epsilon ,k) \ |\ 
(\Lambda;\Lambda ') = (\Lambda^{+};\Lambda '^{+})\}\ .
\end{equation}
A self-conjugate coset label has a representative $(\Lambda
;\Lambda')$ with self-conjugate weights $\Lambda =\Lambda^{+}$,
$\Lambda '=\Lambda'^{+}$. The corresponding vector $d_{\Lambda +\rho
,\Lambda '+\rho}$ is then also self-conjugate, and it therefore has to
lie in the fundamental Weyl chamber or in one if its images under the
action of the invariant subgroup $W^{\omega}\subset W$. For a vector
$d'$ and small enough $\epsilon$, there are no self-conjugate
$d_{\Lambda +\rho ,\Lambda '+\rho}$ in its $\epsilon$-box, unless $d'$
is in one of those Weyl chambers.  Assume therefore that $d'$ is in
such a Weyl chamber, and additionally that $\{d' \}$ (and also the
fractional parts of vectors in its $\epsilon$-box) is contained in the
fundamental affine chamber $C_{0}$ at level $1$. Then the number of
self-conjugate label pairs is given by
\begin{multline}
\left|\{\Lambda =\Lambda^{+},\ \Lambda '=\Lambda '^{+}\ |\ 
| (d'-d_{\Lambda +\rho ,\Lambda '+\rho})_{i}|<\epsilon /2 \} \right|\\
= 
\prod_{i=1}^{m}
\left((k+n+1) 
(\epsilon - | (d')_{i}- (d')_{n-i}|)\Theta (\epsilon - |
(d')_{i}- (d')_{n-i}|)  \right) + \dotsb \ ,
\end{multline}
where $n=2m+1$. 
We again have to ask whether for a given $d$ there are several such
representatives $d'$ on its Weyl orbit. If we adapt the arguments
around eqs.\ \eqref{app:firstw} and~\eqref{app:secondw} to our
situation, we see that this is not the case, because 
the self-conjugate sublattice of the weight lattice agrees with the
self-conjugate sublattice of the root lattice. The cardinality 
$|N^{\omega}(d,\epsilon ,k)|$ is therefore determined by the volume of the
self-conjugate part of the $\epsilon$-box around $d$,
\begin{equation}\label{app:Nomega}
\left| N^{\omega} (d,\epsilon ,k)\right|
=\prod_{i=1}^{m}
\left((k+n+1) 
(\epsilon - | (d)_{i}- (d)_{n-i}|)\Theta (\epsilon - |
(d)_{i}- (d)_{n-i}|)  \right) + \dotsb \ .
\end{equation}

\section{Modular S-matrices}

\subsection{Untwisted coset S-matrix}

The modular S-matrix for the diagonal coset model $\frac{SU (n)_{k}\times SU
(n)_{1}}{SU (n)_{k+1}}$ is given by 
\begin{equation}
S_{(\Lambda,\lambda;\Lambda') (L,l;L')} = n
S^{(n,k)}_{\Lambda L}\overline{S^{(n,k+1)}_{\Lambda' L'}}
S^{(n,1)}_{\lambda l} \ .
\end{equation}
Here, $S^{(n,k)}$ denotes the S-matrix of the affine Lie
algebra $\widehat{sl (n)}_{k}$, which can be written as (see e.g.\ \cite{FrancescoCFT})
\begin{equation}
S^{(n,k)}_{\Lambda L} = i^{n (n-1)/2} n^{-1/2} (k+n)^{-(n-1)/2}
\sum_{w\in W} \epsilon (w) e^{-2\pi i\frac{\left(w (\Lambda
+\rho),L+\rho \right)}{k+n}} \ ,
\end{equation}
with $W$ being the Weyl group and $\rho$ the Weyl vector of $sl (n)$.
Therefore the coset S-matrix can be expressed as
\begin{align}
S_{(\Lambda,\lambda;\Lambda') (L,l;L')} & = \left( (k+n) (k+n+1)\right)^{-(n-1)/2}
S^{(n,1)}_{\lambda l} \nonumber\\
& \quad \times \sum_{w,w'\in W}\epsilon (w)\epsilon (w')
e^{-2\pi i\frac{\left(w (\Lambda
+\rho),L+\rho \right)}{k+n}}
e^{2\pi i\frac{\left(w' (\Lambda'
+\rho),L'+\rho \right)}{k+n+1}} \ .
\label{app:cosetSmatrix}
\end{align}
The coset S-matrix only depends on $\Lambda$ and $\Lambda '$ via their
combination 
\begin{equation}
d_{\Lambda +\rho ,\Lambda '+\rho} = (\Lambda +\rho) - t (\Lambda
'+\rho ) \ ,
\end{equation}
where $t=\frac{k+n}{k+n+1}$. To see this we rewrite 
\begin{equation}
\frac{\Lambda +\rho}{k+n} = t^{-1}d_{\Lambda +\rho ,\Lambda '+\rho} - (\Lambda -\Lambda ')
\end{equation}
and
\begin{equation}
\frac{\Lambda '+\rho}{k+n+1} = d_{\Lambda +\rho ,\Lambda '+\rho} -
(\Lambda -\Lambda ')\ .
\end{equation}
Inserting this into~\eqref{app:cosetSmatrix}, we obtain
\begin{align}
S_{(\Lambda,\lambda;\Lambda') (L,l;L')} & = \left( (k+n) (k+n+1)\right)^{- (n-1)/2}
S^{(n,1)}_{\lambda l} \sum_{w,w'\in W} \epsilon (ww')
e^{-2\pi it^{-1}\left(w (d_{\Lambda +\rho ,\Lambda '+\rho}),L+\rho
\right)}\nonumber\\
&\qquad \times e^{2\pi i\left(w' (d_{\Lambda +\rho,\Lambda '+\rho}),L'+\rho \right)}
e^{2\pi i\left(w (\Lambda
-\Lambda '),L+\rho \right)}
e^{-2\pi i\left(w' (\Lambda
-\Lambda '),L'+\rho \right)} \ .
\end{align}
In the last two exponentials we can replace $w (\Lambda -\Lambda ')$
by $\Lambda -\Lambda '$ (and similarly for $w'$):
$w (\Lambda -\Lambda ')$ differs from $\Lambda -\Lambda'$ 
by an element of the root lattice $L_{R}$, whose scalar product with an
integral weight gives an integer leading to a trivial phase in the
exponential. We thus arrive at
\begin{align}
S_{(\Lambda,\lambda;\Lambda') (L,l;L')} &= \left((k+n) (k+n+1)\right)^{- (n-1)/2}
e^{2\pi i\left(\Lambda-\Lambda ',L-L' \right)}
S^{(n,1)}_{\lambda l} \nonumber\\
&\quad \times \sum_{w,w'\in W} \epsilon (ww')
e^{-2\pi it^{-1}\left(w (d_{\Lambda +\rho ,\Lambda '+\rho}),L+\rho \right)}
e^{2\pi i\left(w' (d_{\Lambda +\rho,\Lambda '+\rho}),L'+\rho \right)}
\ . \label{app:cosetS2}
\end{align}
The S-matrix $S^{(n,1)}$ at level $1$ is very simple, because for
$sl(n)$ all dominant integral weights at level $1$ correspond to
simple currents~\cite{Schellekens:1989am}. We find
\begin{equation}
S^{(n,1)}_{\lambda l} = e^{-2\pi i (\lambda ,l)}S^{(n,1)}_{00} = 
e^{-2\pi i (\lambda ,l)} n^{-1/2} \ .
\end{equation}
From the selection rules~\eqref{selrule} we know that $\Lambda
-\Lambda '+\lambda$ and $L-L'+l$ are in the root lattice
$L_{R}$. Therefore the phases in front of the sum
in~\eqref{app:cosetS2} cancel and we find
\begin{align}
S_{(\Lambda,\lambda;\Lambda') (L,l;L')} &= n^{-1/2}
\left( (k+n) (k+n+1)\right)^{- (n-1)/2}\nonumber\\
& \quad \times \sum_{w,w'\in W} \epsilon (ww')
e^{-2\pi it^{-1}\left(w (d_{\Lambda +\rho ,\Lambda '+\rho}),L+\rho \right)}
e^{2\pi i\left(w' (d_{\Lambda +\rho,\Lambda '+\rho}),L'+\rho \right)}\ .
\label{app:cosetSfinal}
\end{align}

\subsection{Twisted S-matrix}
\label{sec:apptwistedS}
Twisted S-matrices describe the behaviour of characters of twisted
affine Lie algebras under modular transformation~\cite{Kac:1990}. We
are here interested in the case of
$sl(2m+1)$. In~\cite{Gaberdiel:2002qa} (see also~\cite{Fuchs:1996zr})
it was observed that the twisted S-matrix of $sl(2m+1)$ is related to
the untwisted S-matrix of $so(2m+1)$ at level $k+2$, and to the
untwisted S-matrix of $sp(2m)$ at level $(k-1)/2$ (for odd level
$k$). Here we want to express the twisted S-matrix in terms of the
untwisted S-matrix of $sp(2m)$ at level $k+m$. Our starting point is
the determinant formula that can be found e.g.\
in~\cite{Gaberdiel:2002qa}. We label the twisted representations by a
symmetric label $L= (L_{1},\dotsc ,L_{m},L_{m},\dotsc ,L_{1})$. For a
twisted label $L$ and a symmetric weight $\Lambda = (\Lambda_{1},\dotsc
,\Lambda_{m},\Lambda_{m},\dotsc ,\Lambda_{1})$ the entry of the
twisted S-matrix is given by
\begin{equation}\label{app:detformtwist}
\psi_{L\Lambda}^{(2m+1,k)} = (-1)^{\frac{m(m-1)}{2}} \frac{2^{m}}{(k+2m+1)^{m/2}}
\det \left[\sin \left(\frac{2\pi L[i]\Lambda [j]}{k+2m+1} \right)
\right]_{1\leq  i,j\leq m} \ ,
\end{equation}
where 
\begin{equation}
L[i] = m+1 - i +\sum_{j=i}^{m}L_{j} \ ,
\end{equation}
and similarly for $\Lambda [j]$. For the S-matrix of $sp(2m)$ at level
$k+m$, the determinant formula is~\cite{Kac:1988tf}
\begin{equation}\label{app:detformsp}
\hat{S}^{(2m,k+m)}_{\hat{L}\hat{\Lambda}} = (-1)^{\frac{m(m-1)}{2}}
\frac{2^{m/2}}{(k+2m+1)^{m/2}} 
\det \left[\sin \left(\frac{\pi \hat{L}[i]\hat{\Lambda}[j]}{k+2m+1} 
\right) \right]_{1\leq i,j\leq m} \ ,
\end{equation}
where $\hat{L}$ and $\hat{\Lambda}$ are $m$-tuples labelling $sp (2m)$
weights. The two determinants in~\eqref{app:detformtwist} and~\eqref{app:detformsp} are
very similar, and we find
\begin{equation}\label{app:reltwistsp}
\psi^{(2m+1,k)}_{L\Lambda} = 2^{m/2} \hat{S}^{(2m,k+m)}_{\hat{L},
2\hat{\Lambda}+\hat{\rho}} \ .
\end{equation}
Here, the hat $(\,\hat{}\,)$ denotes the map that sends a symmetric $sl
(2m+1)$ weight to a $sp (2m)$-weight,
\begin{equation}
\hat{}\,  : \Lambda = (\Lambda_{1},\dotsc ,\Lambda_{m},\Lambda_{m},\dotsc
,\Lambda_{1}) \mapsto \hat{\Lambda}=(\Lambda_{1},\dotsc ,\Lambda_{m})
\ .
\end{equation}
Under this map the Weyl vector $\rho$ of $sl (2m+1)$ is mapped to the
Weyl vector $\hat{\rho}$ of $sp (2m)$.

Using standard expressions for untwisted modular S-matrices (see e.g.\
\cite{FrancescoCFT}), we can rewrite~\eqref{app:reltwistsp} as
\begin{equation}\label{app:twistS}
\psi_{L\Lambda}^{(2m+1,k)}=  i^{m^{2}} (k+2m+1)^{-m/2} \sum_{w\in\hat{W}} \epsilon (w)
e^{-2\pi i\frac{\left( w(\hat{L}+\hat{\rho}),2\hat{\Lambda}+2\hat{\rho} \right)}{k+2m+1}} \ ,
\end{equation}
where $\hat{W}$ is the Weyl group of $sp(2m)$. The scalar product
appearing in the exponential is the standard quadratic form of the
$sp(2m)$ algebra. It is related to the quadratic form on the $sl
(2m+1)$ weight space by~\cite{Fuchs:1996zr}
\begin{equation}\label{app:quadform}
(L,\Lambda)_{sl (2m+1)} = 4 (\hat{L},\hat{\Lambda})_{sp (2m)} \ .
\end{equation}
The action of the Weyl group $\hat{W}$ on $sp (2m)$ weights induces an
action on symmetric $sl (2m+1)$ weights which corresponds precisely to
the action of the subgroup $W^{\omega}\subset W$ of all $sl (2m+1)$ Weyl
transformations that map symmetric weights to symmetric
weights. Therefore we can rewrite the twisted S-matrix as
\begin{equation}\label{app:twistSfinal}
\psi_{L\Lambda}^{(2m+1,k)}=  i^{m^{2}} (k+2m+1)^{-m/2} \sum_{w\in
W^{\omega}} \epsilon (w)
e^{-\pi i\frac{\left( w(L+\rho),\Lambda+\rho \right)}{k+2m+1}} \ ,
\end{equation}
which coincides with the expression given in~\cite{Birke:1999ik}. 

When we discuss boundary conditions in the limit of minimal models, we
want to make sense of symmetric boundary labels $L$ that are outside of the
usual range and do not satisfy $\sum_{i=1}^{n-1}L_{i}\leq k$. The
formulae~\eqref{app:twistS} and~\eqref{app:twistSfinal} can also be
applied for those labels $L$. In particular, in the $sp (2m)$
language, the formula~\eqref{app:twistS} giving the twisted S-matrix
in terms of $\hat{L}$ is invariant under
a shifted Weyl reflection by $w\in \hat{W}$ (up to a sign), and under
translations by $(k+n)/2$-multiples of co-root vectors. These
transformations can be interpreted as the shifted action of the affine
Weyl group at level $(k-1)/2$ (this action also makes sense for even
$k$), and these transformations can be used to map any label to some $\hat{L}$
satisfying $\sum_{i=1}^{m}\hat{L}_{i}\leq k/2$. In the $sl (2m+1)$
language this also has a natural interpretation. The lattice spanned
by half the coroot vectors of $sp (2m)$ coincides with the weight
lattice of $sp (2m)$. Translations of $\hat{L}$ by the weight lattice
of $sp (2m)$ correspond to translations of $L$ by the symmetric
(self-conjugate) part of the weight lattice of $sl (2m+1)$. This in
turn coincides with the symmetric part of the root lattice of $sl
(2m+1)$. Therefore we can use the symmetric part of the affine Weyl
group to bring any symmetric label to some $L$ lying in the usual range.

\subsection{Twisted coset S-matrix}

The twisted coset S-matrix for the $sl(2m+1)$ diagonal coset model is
given by
\begin{equation}
\psi_{(L,L')(\Lambda ,\Lambda ')} =
\psi^{(2m+1,k)}_{L\Lambda} \psi^{(2m+1,1)}_{00}
\overline{\psi}^{(2m+1,k+1)}_{L'\Lambda '} \ ,
\end{equation}
where $\Lambda =\Lambda^{+}$ and $\Lambda '=\Lambda '^{+}$ are
self-conjugate labels. The twisted S-matrix for the level~$1$ part is
trivial, $\psi^{(2m+1,1)}_{00}=1$, and can be omitted.

\noindent Using~\eqref{app:twistS} the twisted S-matrix takes the form
\begin{align}
\psi_{(L,L')(\Lambda ,\Lambda ')} & =
\left( (k+2m+1) (k+2m+2)\right)^{-m/2} \nonumber\\
& \quad \times \sum_{w,w'\in \hat{W}} \epsilon (ww')
e^{-2\pi i \frac{\left(w(\hat{L}+\hat{\rho} ),2\hat{\Lambda}+2\hat{\rho}  \right)}{k+2m+1}}
e^{2\pi i\frac{\left(w'(\hat{L}'+\hat{\rho}),2\hat{\Lambda}'+2\hat{\rho} \right)}{k+2m+2}} 
\ .
\end{align}
We rewrite
\begin{equation}
\frac{\hat{\Lambda}+\hat{\rho} }{k+2m+1} = 
t^{-1}\left((\hat{\Lambda}+\hat{\rho})-t(\hat{\Lambda}'+\hat{\rho} ) \right) 
-(\hat{\Lambda}-\hat{\Lambda}') 
\end{equation}
and 
\begin{equation}
\frac{\hat{\Lambda}'+\rho}{k+2m+2} = 
(\hat{\Lambda}+\rho)-t(\hat{\Lambda}'+\rho )
-(\hat{\Lambda}-\hat{\Lambda}') \ ,
\end{equation}
where $t=\frac{k+2m+1}{k+2m+2}$. For the combination of
$\hat{\Lambda}$ and $\hat{\Lambda}'$ we introduce the notation
\begin{equation}
\hat{d}_{\Lambda +\rho ,\Lambda '+\rho '} =
(\hat{\Lambda}+\hat{\rho})-t (\hat{\Lambda}'+\hat{\rho}) \ .
\end{equation}
This allows us to express the S-matrix as
\begin{align}
\psi_{(L,L')(\Lambda ,\Lambda ')} &=
\left( (k+2m+1) (k+2m+2)\right)^{-m/2} \nonumber\\
& \quad \times \sum_{w,w'\in \hat{W}} \epsilon (ww')
e^{-2\pi i\left(t^{-1}w(\hat{L}+\hat{\rho} )-w'(\hat{L}'+\hat{\rho} ), 
2\hat{d}_{\Lambda +\rho ,\Lambda'+\rho'} \right) }\nonumber\\
&\qquad \times e^{2\pi i \left(w(\hat{L}+\hat{\rho} )
-w'(\hat{L}'+\hat{\rho}),2 (\hat{\Lambda}-\hat{\Lambda}') \right)} \ .
\end{align}
The phase in the last line is trivial: the quadratic form on the
weight lattice of $sp(2m)$ takes values in $\frac{1}{2}\mathbb{Z}$,
and $2 (\hat{\Lambda}-\hat{\Lambda}')$ is an even weight vector, therefore
it has integer scalar product with any vector in the weight lattice.
Finally we express everything in terms of symmetric $sl (2m+1)$ labels
(similar to~\eqref{app:twistSfinal}) and we obtain
\begin{align}
\psi_{(L,L')(\Lambda ,\Lambda ')} &=
\left( (k+2m+1) (k+2m+2)\right)^{-m/2} \nonumber\\
& \quad \times \sum_{w,w'\in W^{\omega}} \epsilon (ww')
e^{-\pi i\left(t^{-1}w(L+\rho)-w'(L'+\rho), 
d_{\Lambda +\rho ,\Lambda'+\rho'} \right) } \ .
\label{app:cosettwistSfinal}
\end{align}

\end{appendix}

\bibliographystyle{mystyle4}
\bibliography{references}

\begin{thebibliography}{10}

\bibitem{Belavin:1984vu}
A.~A. Belavin, A.~M. Polyakov, A.~B. Zamolodchikov, {\em {Infinite conformal
  symmetry in two-dimensional quantum field theory}\/}, Nucl. Phys. {\bf B241}
  (1984) 333

\bibitem{Fuchs:2009iz}
J.~Fuchs, I.~Runkel, C.~Schweigert, {\em {Twenty-five years of two-dimensional
  rational conformal field theory}\/}, J. Math. Phys. {\bf 51} (2010) 015210,
  0910.3145

\bibitem{Runkel:2001ng}
I.~Runkel, G.~M.~T. Watts, {\em A non-rational CFT with $c = 1$ as a limit of
  minimal models\/}, JHEP {\bf 09} (2001) 006, hep-th/0107118

\bibitem{Schomerus:2003vv}
V.~Schomerus, {\em Rolling tachyons from Liouville theory\/}, JHEP {\bf 11}
  (2003) 043, hep-th/0306026

\bibitem{Fredenhagen:2004cj}
S.~Fredenhagen, V.~Schomerus, {\em Boundary Liouville theory at $c = 1$\/},
  JHEP {\bf 05} (2005) 025, hep-th/0409256

\bibitem{Fredenhagen:2007tk}
S.~Fredenhagen, D.~Wellig, {\em {A common limit of super Liouville theory and
  minimal models}\/}, JHEP {\bf 09} (2007) 098, 0706.1650

\bibitem{Roggenkamp:2003qp}
D.~Roggenkamp, K.~Wendland, {\em {Limits and degenerations of unitary conformal
  field theories}\/}, Commun. Math. Phys. {\bf 251} (2004) 589, hep-th/0308143

\bibitem{Dorn:1994xn}
H.~Dorn, H.~J. Otto, {\em Two and three point functions in Liouville theory\/},
  Nucl. Phys. {\bf B429} (1994) 375, hep-th/9403141

\bibitem{Zamolodchikov:1995aa}
A.~B. Zamolodchikov, A.~B. Zamolodchikov, {\em Structure constants and
  conformal bootstrap in Liouville field theory\/}, Nucl. Phys. {\bf B477}
  (1996) 577, hep-th/9506136

\bibitem{Fateev:2000ik}
V.~Fateev, A.~B. Zamolodchikov, A.~B. Zamolodchikov, {\em Boundary Liouville
  field theory. I: Boundary state and boundary two-point function\/}  (2000),
  hep-th/0001012

\bibitem{Teschner:2000md}
J.~Teschner, {\em Remarks on Liouville theory with boundary\/}  (2000),
  hep-th/0009138

\bibitem{Zamolodchikov:2001ah}
A.~B. Zamolodchikov, A.~B. Zamolodchikov, {\em Liouville field theory on a
  pseudosphere\/}  (2001), hep-th/0101152

\bibitem{Hosomichi:2001xc}
K.~Hosomichi, {\em Bulk-boundary propagator in Liouville theory on a disc\/},
  JHEP {\bf 11} (2001) 044, hep-th/0108093

\bibitem{Ponsot:2001ng}
B.~Ponsot, J.~Teschner, {\em Boundary Liouville field theory: Boundary three
  point function\/}, Nucl. Phys. {\bf B622} (2002) 309, hep-th/0110244

\bibitem{Ponsot:2003ss}
B.~Ponsot, {\em {Liouville theory on the pseudosphere: Bulk-boundary structure
  constant}\/}, Phys. Lett. {\bf B588} (2004) 105, hep-th/0309211

\bibitem{Campoleoni:2010zq}
A.~Campoleoni, S.~Fredenhagen, S.~Pfenninger, S.~Theisen, {\em {Asymptotic
  symmetries of three-dimensional gravity coupled to higher-spin fields}\/},
  JHEP {\bf 11} (2010) 007, 1008.4744

\bibitem{Henneaux:2010xg}
M.~Henneaux, S.-J. Rey, {\em {Nonlinear W(infinity) Algebra as Asymptotic
  Symmetry of Three-Dimensional Higher Spin Anti-de Sitter Gravity}\/}  (2010),
  1008.4579

\bibitem{Gaberdiel:2010pz}
M.~R. Gaberdiel, R.~Gopakumar, {\em {An AdS$_3$ Dual for Minimal Model CFTs}\/}
   (2010), 1011.2986

\bibitem{Alday:2009aq}
L.~F. Alday, D.~Gaiotto, Y.~Tachikawa, {\em {Liouville Correlation Functions
  from Four-dimensional Gauge Theories}\/}  (2009), 0906.3219

\bibitem{Wyllard:2009hg}
N.~Wyllard, {\em {A$_{N-1}$ conformal Toda field theory correlation functions
  from conformal N=2 SU(N) quiver gauge theories}\/}, JHEP {\bf 11} (2009) 002,
  0907.2189

\bibitem{Fateev:2007ab}
V.~A. Fateev, A.~V. Litvinov, {\em {Correlation functions in conformal Toda
  field theory I}\/}, JHEP {\bf 11} (2007) 002, 0709.3806

\bibitem{Fateev:2008bm}
V.~A. Fateev, A.~V. Litvinov, {\em {Correlation functions in conformal Toda
  field theory II}\/}, JHEP {\bf 01} (2009) 033, 0810.3020

\bibitem{Drukker:2010jp}
N.~Drukker, D.~Gaiotto, J.~Gomis, {\em {The Virtue of Defects in 4D Gauge
  Theories and 2D CFTs}\/}  (2010), 1003.1112

\bibitem{Fateev:2010za}
V.~Fateev, S.~Ribault, {\em {Conformal Toda theory with a boundary}\/}  (2010),
  1007.1293

\bibitem{Cardy:1989ir}
J.~L. Cardy, {\em Boundary conditions, fusion rules and the {Verlinde}
  formula\/}, Nucl. Phys. {\bf B324} (1989) 581

\bibitem{Fateev:2001mj}
V.~A. Fateev, {\em {Normalization factors, reflection amplitudes and integrable
  systems}\/}  (2001), hep-th/0103014

\bibitem{Bais:1987zk}
F.~A. Bais, P.~Bouwknegt, M.~Surridge, K.~Schoutens, {\em {Coset Construction
  for Extended Virasoro Algebras}\/}, Nucl. Phys. {\bf B304} (1988) 371

\bibitem{Goddard:1986ee}
P.~Goddard, A.~Kent, D.~I. Olive, {\em Unitary Representations of the Virasoro
  and Supervirasoro Algebras\/}, Commun. Math. Phys. {\bf 103} (1986) 105

\bibitem{FrancescoCFT}
P.~D. Francesco, P.~Mathieu, D.~S{\'{e}}n{\'{e}}chal, {\em {Conformal Field
  Theory}\/}, Graduate Texts in Contemporary Physics, Springer, New York (1999)

\bibitem{Birke:1999ik}
L.~Birke, J.~Fuchs, C.~Schweigert, {\em Symmetry breaking boundary conditions
  and {WZW} orbifolds\/}, Adv. Theor. Math. Phys. {\bf 3} (1999) 671,
  hep-th/9905038

\bibitem{Ishikawa:2001zu}
H.~Ishikawa, {\em Boundary states in coset conformal field theories\/}, Nucl.
  Phys. {\bf B629} (2002) 209, hep-th/0111230

\bibitem{Fredenhagen:2003xf}
S.~Fredenhagen, {\em Organizing boundary RG flows\/}, Nucl. Phys. {\bf B660}
  (2003) 436, hep-th/0301229

\bibitem{Caldeira:2004jy}
A.~F. Caldeira, J.~F. Wheater, {\em {Boundary states and broken bulk symmetries
  in W A(r) minimal models}\/}  (2004), hep-th/0404052

\bibitem{Fateev:1987zh}
V.~A. Fateev, S.~L. Lukyanov, {\em {The Models of Two-Dimensional Conformal
  Quantum Field Theory with Z(n) Symmetry}\/}, Int. J. Mod. Phys. {\bf A3}
  (1988) 507

\bibitem{Fredenhagen:2001kw}
S.~Fredenhagen, V.~Schomerus, {\em D-branes in coset models\/}, JHEP {\bf 02}
  (2002) 005, hep-th/0111189

\bibitem{Recknagel:2000ri}
A.~Recknagel, D.~Roggenkamp, V.~Schomerus, {\em On relevant boundary
  perturbations of unitary minimal models\/}, Nucl. Phys. {\bf B588} (2000)
  552, hep-th/0003110

\bibitem{Graham:2001tg}
K.~Graham, I.~Runkel, G.~M.~T. Watts, {\em Minimal model boundary flows and $c
  = 1$ {CFT}\/}, Nucl. Phys. {\bf B608} (2001) 527, hep-th/0101187

\bibitem{Alekseev:2002rj}
A.~Y. Alekseev, S.~Fredenhagen, T.~Quella, V.~Schomerus, {\em Non-commutative
  gauge theory of twisted {D}-branes\/}, Nucl. Phys. {\bf B646} (2002) 127,
  hep-th/0205123

\bibitem{Gaberdiel:2003kv}
M.~R. Gaberdiel, T.~Gannon, {\em The charges of a twisted brane\/}, JHEP {\bf
  01} (2004) 018, hep-th/0311242

\bibitem{Fredenhagen:thesis}
S.~Fredenhagen, {\em D-brane dynamics in curved backgrounds\/}, Ph.D. thesis,
  Humboldt University, Berlin (2002),
  (http://edoc.hu-berlin.de/abstract.php3/\\
  dissertationen/fredenhagen-stefan-2002-09-16)

\bibitem{Tachikawa:2010vg}
Y.~Tachikawa, {\em {N=2 S-duality via Outer-automorphism Twists}\/}  (2010),
  1009.0339

\bibitem{Lukyanov:1990tf}
S.~L. Lukyanov, V.~A. Fateev, {\em Physics reviews: Additional symmetries and
  exactly soluble models in two-dimensional conformal field theory\/} Chur,
  Switzerland: Harwood (1990) 117 p. (Soviet Scientific Reviews A, Physics:
  15.2)

\bibitem{Kiritsis:2010xc}
E.~Kiritsis, V.~Niarchos, {\em {Large-N limits of 2d CFTs, Quivers and AdS$_3$
  duals}\/}  (2010), 1011.5900

\bibitem{Schellekens:1989am}
A.~N. Schellekens, S.~Yankielowicz, {\em Extended Chiral Algebras and Modular
  Invariant Partition Functions\/}, Nucl. Phys. {\bf B327} (1989) 673

\bibitem{Kac:1990}
V.~G. Kac, {\em Infinite Dimensional {Lie} Algebras\/}, Cambridge University
  Press, Cambridge (1990)

\bibitem{Gaberdiel:2002qa}
M.~R. Gaberdiel, T.~Gannon, {\em Boundary states for {WZW} models\/}, Nucl.
  Phys. {\bf B639} (2002) 471, hep-th/0202067

\bibitem{Fuchs:1996zr}
J.~Fuchs, B.~Schellekens, C.~Schweigert, {\em From {Dynkin} diagram symmetries
  to fixed point structures\/}, Commun. Math. Phys. {\bf 180} (1996) 39,
  hep-th/9506135

\bibitem{Kac:1988tf}
V.~G. Kac, M.~Wakimoto, {\em {Modular and conformal invariance constraints in
  representation theory of affine algebras}\/}, Adv. Math. {\bf 70} (1988) 156

\end{thebibliography}

\end{document}